\documentclass[letterpaper,galley,usenatbib]{mn2e}
\usepackage{times}
\usepackage{dcolumn}
\usepackage{myaasmacros}
\usepackage{mathrsfs}
\usepackage{amsmath}
\usepackage{array}
\usepackage{multirow}
\usepackage{amssymb}
\usepackage{amsfonts}
\usepackage{graphicx}
\usepackage{wasysym}
\usepackage{wrapfig}
\usepackage{dblfloatfix}
\usepackage{fixltx2e}
\usepackage{color}

\newcommand{\gasLow}{{\textit{gasLow}}}
\newcommand{\gasHigh}{{\textit{gasHigh}}}
\newcommand{\gasHighd}{{\textit{gasHigh\_d}}}
\newcommand{\gasLowfb}{{\textit{gasLow\_fb}}}
\newcommand{\gasHighfb}{{\textit{gasHigh\_fb}}}
\newcommand{\gasHighdfb}{{\textit{gasHigh\_d\_fb}}}

\newcommand{\pspher}{{\textit{p\_spher}}}
\newcommand{\pdens}{{\textit{p\_dens}}}
\newcommand{\pressure}{{\textit{pressure}}}

\newcommand{\highRes}{\textit{highRes}}
\newcommand{\lowRes}{\textit{lowRes}}

\newcommand{\Msun}{\ensuremath{\mathrm{M}_\odot }}

\newcommand{\twoFone}[1]{\ensuremath{{}_2F_1}}
\renewcommand{\vec}[1]{\ensuremath{\boldsymbol{#1}}}
\definecolor{grey}{rgb}{0.4,0.6,0.6}
\definecolor{darkgreen}{rgb}{0.0,0.7,0.0}
\definecolor{darkblue}{rgb}{0.0,0.3,0.7}
\definecolor{darkred}{rgb}{0.5,0.,0.}

\newcommand\plotone[1]{%
 \centering
 \leavevmode
 \includegraphics[width={\columnwidth}]{#1}%
}%
\newcommand\plotoneF[1]{%
 \centering
 \leavevmode
 \includegraphics[width={\textwidth}]{#1}%
}%

\title[\Pressure]{External pressure-triggering of star formation in  a disc galaxy: a template for positive feedback}
\author[R. Bieri  et al. ]{
\parbox[t]{\textwidth}{
Rebekka Bieri$^{1}$\thanks{E-mail: bieri@iap.fr},
Yohan Dubois$^1$, Joseph Silk$^{1,2,3,4}$, Gary A. Mamon$^1$ \\
and Volker Gaibler$^{5}$}
\vspace*{6pt} \\
$^{1}$ Institut d'Astrophysique de Paris (UMR 7095: CNRS \& UPMC -- Sorbonne
Universit\'es), 98 bis bd Arago, F-75014 Paris, France\\ 
$^2$ Laboratoire AIM-Paris-Saclay, CEA/DSM/IRFU, CNRS, Univ. Paris VII, F-91191 Gif-sur-Yvette, France\\
$^{3}$ Department of Physics and Astronomy, The Johns Hopkins University Homewood Campus, Baltimore, MD 21218, USA\\
$^{4}$ BIPAC, Department of Physics, University of Oxford, Keble Road, Oxford
OX1 3RH\\
$^5$ Institut f\"ur Theoretische Astrophysik, Universit\"at Heidelberg, Albert-Ueberle-Str 2, 
D--69120 Heidelberg, Germany\\
}
\date{Accepted . Received ; in original form }

\begin{document}
\maketitle

\begin{abstract}
Feedback from active galactic nuclei (AGN) has often been invoked  both in
simulations and in interpreting observations for regulating star formation and
quenching cooling flows in massive galaxies.  AGN activity can, however, also
over-pressurise the dense star-forming regions of galaxies and thus enhance
star formation, leading to a positive feedback effect. To understand this
pressurisation better, we investigate the effect of an ambient external
pressure on gas fragmentation and triggering of starburst activity by means of
hydrodynamical simulations.  We find that moderate levels of over-pressurisation
of the galaxy boost the global star formation rate of the galaxy by an order of
magnitude, turn stable discs unstable, and lead to significant fragmentation of
the gas content of the galaxy, similar to what is  observed in high redshift
galaxies.
\end{abstract}


\section{Introduction}
\label{sec:intro} 

Supermassive black holes are found at the centers of most, if not all, massive
galaxies (e.g., \citealp{Magorrian1998}; \citealp{Hu2008};
\citealp{Kormendy+2011}).  Throughout cosmic history, they are thought to play
an important role in regulating the baryonic mass content of massive galaxies
through {\it feedback} from Active Galactic Nuclei (AGN) by releasing a
fraction of the rest-mass accreted energy back into the galactic gas and
altering the star formation rate (SFR) in the galaxy.

AGN can exert either {\it negative} or {\it positive} feedback on their
surroundings.  The former describes cases where the AGN inhibits star formation
by heating and dispersing the gas in the galaxy, while the latter describes the
possibility that an AGN may trigger star formation.  Negative AGN feedback can
operate in \textit{quasar-mode} from radiation at high accretion rates, or
\textit{radio-mode} from AGN jets at predominantly low accretion
rates~\citep{churazovetal05,russelletela13}. It is still unclear how
efficiently AGN feedback delivers energy (through heating, e.g.,
\citealp{Silk+Rees1998}) and momentum (through physical pushing,
\citealp{King2003}) to the galaxy's gas and what mode of feedback dominates.
Both semi-analytical \citep[e.g.,][]{Croton+2006, Bower+2006} and
hydrodynamical cosmological simulations \citep[e.g.,][]{DiMatteo+Springel+2005,
DiMatteo+2008, Sijacki+2007, BoothSchaye2009, duboisetal2010} have shown that
negative AGN feedback is an important ingredient in the formation and evolution
of massive galaxies, in particular in shaping the observed high-end tail of the
galaxy mass function, and the low SFRs in massive galaxies.  Moreover,
observations show that cooling flows in the hot circumgalactic and intracluster
media can be suppressed by the energy transferred by AGN jets
\citep{Birzan2004, Dunn+Fabian+2005}, again negatively impacting star
formation. 

Although AGN feedback has been extensively studied in observations and through
cosmological simulations, the impact on the host galaxy and the precise
mechanism of the communication of the AGN with the galaxy's interstellar medium
(ISM) is far from being understood.  It is not clear why jet feedback, which is
thought to heat cold gas, should have a similar effect on the multi-phase ISM.
It has been argued that a jet that propagates through an inhomogeneous ISM may
also trigger or enhance star formation in a galaxy (i.e., positive
feedback). \citet{Begelman+Cioffi1989} and \citet{Rees1989} proposed that the
radio jet activity triggers star formation and might serve as an explanation
for the alignment of radio and optical structures in high redshift radio
galaxies. Radio jet-induced star formation has also been considered as a
source powering luminous starbursts \citep{Silk2005}.
\citet{Ishibashi+Fabian2012} provide a theoretical framework linking AGN
feedback triggering of star formation in the host galaxy to the oversized
evolution of massive galaxies over cosmic time.  Furthermore, negative and
positive feedback are not necessarily contradictory
\citep{Silk2013,Zubovas+2013,Zinn+2013,Cresci+2015}: AGN activity may both
quench and induce star formation in different parts of the host galaxy and
on different time-scales.

Observationally, this positive feedback scenario is directly supported by only
a few local \citep{Croft+2006, Inskip+2008, salomeetal15} and high redshift
\citep{Dey+1997, Bicknell+Sutherland+2000, rauchetal13} observations. There
are, however, also indirect links between jets and star formation which suggest
possible positive feedback from AGN \citep{Klamer+2004, McCarthy1991,
McCarthy1993, Balmaverde+2008, Podigachoski2015, swinbanketal15}.

More recently, high resolution hydrodynamical simulations of a jet including a
multi-phase ISM have become feasible  \citep{Sutherland+Bicknell2007,
Antonuccio-Delogu+Silk2008, Antonuccio-Delogu+Silk2010, Wagner+Bicknell2011,
Gaibler+2012}.  These studies have shown that a clumpy interstellar structure
results in a different interaction between the jet and the gas than was assumed
from simulations with a homogeneous ISM. It can be generally noted that an
inhomogeneous ISM affects not only the jet evolution, but also the morphology
of the host galaxy itself.  Simulations by \citet{Tortora+2009} extended the
studies and generalised the simulations of \citet{Antonuccio-Delogu+Silk2008}
studying the interaction of a powerful jet in 2D, two-phase ISM.
\citet{Tortora+2009} have shown that star formation can initially be slightly
increased (10--20 per cent) followed by a much stronger quenching (more than 50
per cent) within a time-scale of a few million years.  They argue that the
rapid decrease of the SFR after its initial enhancement is a consequence of
both the high temperatures as well as the reduced cloud mass once the jet
cocoon has propagated within the medium.  Kelvin-Helmholtz instabilities reduce
the mass of the clouds and, assuming a Schmidt-Kennicutt law, thereby reduce
the SFR.  It should, however, be noted that the 2D approach
results in a very different temperature and pressure evolution as compared to a
3D simulation.  \citet{Wagner+Bicknell2011} studied the interaction of a
relativistic jet interacting with a two-phase ISM at the galaxy's center with a
resolution of one kiloparsec. They found that the transfer of energy and
momentum to the ISM may inhibit star formation through the dispersal of gas,
but their simulations do not contain a star-formation model.  It could be
argued, though, that because of the short ($\la 1$~Myr) simulation timescale,
the impact of cooling is very weak and they therefore might underestimate the
SFR.

\citet{Gaibler+2012} simulated a powerful AGN jet within a massive gaseous,
clumpy disc (however they neglected gravity).  They  showed that the jet
activity causes a significant change of the SFR by enhancing the formation of
stars, with  inside-out propagation in the galaxy (see~\citealp{duganetal14}).
Their simulations show the formation of a blast wave in the central
region of the disc possibly dominating the early evolution between the jet and
the galaxy. The blast wave results in the formation of a cavity in the disc
center pushing the gas outwards and compressing the gas within the disc at the
cavity boundary,  generating rings of compressed gas within the disc. At later
times, the ISM of the disc is pressurised by the bow shock enclosing the entire
disc. This pressurisation is both due to the ram pressure from the backflow but
also the thermal pressure of the cocoon.  It is expected that the thermal
pressure dominates somewhat as the turbulence is measured to be in the subsonic
or transonic regime. 

Although the physical understanding of star formation is still limited and
debated \citep{Padoan+Nordlund2011}, it can be assumed that a pressurised disc
can trigger gravitational instabilities, compress the galaxy's clouds, and push
the densities within the disc above the critical density for star formation,
thus resulting in an increased SFR.  
This picture is further supported by a few observations of well-resolved star-forming
molecular clouds~\citep{ketoetal05, rosolowsky&blitz05} and by detailed simulation 
of the ISM~\citep[e.g.][]{slyzetal05, zubovasetal14}.
 
Motivated by the pressurisation of the disc found by \cite{Gaibler+2012} in
their hydrodynamical simulations of AGN jet feedback, we have investigated the
effects of this extra pressure on a galaxy disc, by running
hydrodynamical simulations with self-gravity, without AGN jets, but with simple
prescriptions for external pressure such as that which may be caused by the jet
cocoon.  In a first study \citep{Bieri+15}, we simulated disc galaxies of
one-tenth the total mass of the Milky Way, varying their initial gas fraction.
We found that with a given level of external pressure, the  disk fragments into
numerous clumps, causing enhanced star formation.  In the present article, we
study the effects of external pressure in more detail, by considering different
geometries and levels of external pressure, as well as studying the effects of
supernova feedback and mass resolution.

In Section~\ref{sec:setup}, we describe our suite of hydrodynamical
simulations.  Our results are presented in Section~\ref{sec:results} and
summarised in Section~\ref{sec:conclusions}.

\section{Simulation Set-up}
\label{sec:setup} 

\subsection{Basic simulation scheme}

\begin{table} 
\caption{Galaxy parameters: scale radius ($r_s$), gas fraction ($f_g$), total
stellar mass ($M_\mathrm{*}$), and total gas mass in the disc ($M_{\mathrm{gas}}$)}
\begin{center}
\begin{tabular}{lcccc} 
\hline Identifier & $r_s$ & $f_g$ & $M_{\mathrm{*}}$ & $M_{\mathrm{gas}}$ \\ 
 & [kpc] & $[\%]$ & [$10^9$ $\mathrm{M}_{\odot}$]  & [$10^9$ $\mathrm{M}_{\odot}$] \\ \hline
gasLow     & 3.4 & 10 & 8.1 & 0.9 \\
gasHigh    & 3.4 & 50 & 4.6 & 4.4 \\\hline 
\end{tabular} 
\end{center}
\label{tab:init1}
\end{table}
 
Our simulations begin with a galaxy made of a disc of gas and stars, a stellar
bulge and a dark matter (DM) halo. We allow this galaxy to relax to an
equilibrium configuration (with a reasonable disc thickness) over the rotation
time of the disc at its half-mass radius. This first phase is performed without
gas cooling, or star formation or feedback,  in order to evacuate spurious
waves emitted from the imperfect equilibrium of the initial conditions. After
this first relaxation phase, we turn on the external pressure, gas cooling,
star formation, and also feedback from supernovae (SNe), as described below.

The initial condition method introduced by \citet{Springel+Hernquist2005} is
used to generate the DM particles with an NFW~\citep{NFW97} density profile and
a concentration parameter of $c=10$.  The virial velocity of the DM particles
is set to be $v_{200} = 70 \,�\rm km\, s^{-1}$, which corresponds to a virial
radius of $R_{200} \approx 96 \,\rm kpc$ and a virial mass of $M_{200} \approx
1.1 \times 10^{11}\, \rm \Msun$. A Hubble constant of $H_0 = 73\,�\rm km\,
s^{-1}\, Mpc^{-1}$ is assumed. The star particles as well as the gas are
distributed in an exponential disc with a scale length of $3.44\,�\rm kpc$ and
scale height $0.2\, \rm kpc$, and a spherical, non rotating bulge with a
Hernquist profile \citep{Hernquist1990} of scale radius $0.2\, \rm kpc$. We use
$10^6$ DM particles with a mass resolution of $1.23 \times 10^{5}\, \Msun$ to
sample the dark matter halo, and $5.625\times10^5$ star particles sampling the
disc of which $6.25\times10^4$ star particles are used to sample the bulge. The
stellar mass resolution is $1.57 \times 10^{4}\, \rm \Msun$ for the 10\% gas
fraction simulation (hereafter, \gasLow) whereas for the 50\% gas fraction
simulation (hereafter, \gasHigh) the mass resolution is $8.73 \times 10^{3}\,
\rm \Msun$. The relevant galaxy parameters are shown in Table~\ref{tab:init1}.
 
The simulations are run with the {\sc Ramses} adaptive mesh refinement code
\citep{Teyssier2002}. Particles motions are evolved through the gravitational
force with an adaptive particle mesh solver using a cloud-in-cell
interpolation, together with the mass contribution of the gas component.  The
evolution of the gas is followed with a second-order unsplit Godunov scheme.
We use the HLLC Riemann solver~\citep{Toro+1994} with MinMod total variation
diminishing scheme to reconstruct the interpolated variables from their
cell-centred values.  The box size is $655\, \rm kpc$ with a coarse level of 7,
and a maximum level of 14 corresponding to a $\Delta x=40\, \rm pc$ minimum
cell size for most of the simulations.  For convergence studies, we perform a
higher resolution run with a spatial resolution of $\Delta x=10\,�\rm pc$
(maximum level of refinement 16).  The refinement is triggered with a
quasi-Lagrangian criterion: if the gas mass within a cell is larger than $8
\times 10^{7}\, \rm \Msun$ or if more than 8 DM particles are within the cell a
new refinement level is triggered.

The circumgalactic medium is modelled with a constant hydrogen number density
of $n_{\rm CGM}=10^{-3}\,\rm H\, cm^{-3}$.  The pressure and temperature
profiles outside the disc are calculated assuming spherical hydrostatic
equilibrium. For the relaxation phase, the simulations are run for one rotation
period of the half-baryonic mass radius (5 kpc) of the galaxy, i.e.  $\approx 0.5\,
\rm Gyr$. 

The simulations include sub-grid models for cooling, star formation, as well as
SN feedback in a subset of runs.  The cooling mechanism is that described by
\citet{Sutherland+Dopita1993}, which accounts for H, He, and metal
contributions to gas cooling (assuming a solar chemical composition of the
various metal elements, but with a varying metallicity of the gas).  The disc
is initialised with a uniform solar metallicity. No metals are initially placed
outside the disc. The boundary of the disc is defined using a geometrical
criteria with cylindrical symmetry using the initial disc radius and disc
height.  Metals are passively advected with gas in the simulation and are
modified by individual SNe events with a yield of $0.1$, which also distribute
the metals throughout and outside the galaxy.  In dense and cold regions, gas
is turned into star particles following a Schmidt law:
\begin{equation}
\dot\rho_* = \epsilon_* {\rho_{\rm gas} \over t_{\rm ff}} \quad \hbox{if}
\quad n_{\rm gas} > n_0 \ ,
\label{eq:sfr}
\end{equation}
where $\dot \rho_*$ is the star formation rate density, $\rho_{\rm gas}$ is the
gas mass density, $\epsilon_*=0.01$ is the star formation efficiency, $t_{\rm
ff}$ is the local gas free-fall time, and $n_{\rm gas}$ and $n_{0}=14\,\rm H\,
cm^{-3}$ for $\Delta x=40\, \rm pc$ ($n_{0}=224\,\rm H\,  cm^{-3}$ for $\Delta
x=10\, \rm pc$) are the local H number density and H number density threshold
for star formation respectively.  The Schmidt law is used to draw a probability
to form a star with a stellar mass of $m_*=\rho_0\Delta x^3\simeq3\times 10^4
\,�\rm M_\odot$ for the low resolution (\lowRes) runs and a stellar mass of
$m_*\simeq7\times 10^3 \,�\rm M_\odot$ for the high resolution (\highRes)
runs~\citep{Rasera+Teyssier2006}.  The gas temperature in high gas density
regions ($n_{\rm gas}>n_0$) is artificially enhanced by a polytropic equation
of state $T=T_0(n_{\rm gas}/n_0)^{\kappa-1}$, where $\kappa=2$ is the
polytropic index, and $T_0=270\, \rm K$ for the low and high resolution runs.
It is chosen in order to get a constant Jeans length resolved with at least 4
cells. This artificial polytropic equation of state is used to prevent the
catastrophic and artificial collapse of the self-gravitating
gas~\citep{trueloveetal97}. 

We account for the mass and energy release from type II SNe.  The energy
injection, which is purely thermal, corresponds to  
\begin{equation} 
E _{\rm SN} = \eta_{\text{SN}}\, {m_* \over {\rm M_\odot}} 10^{50} \; \text{erg} \, ,
\end{equation} 
where $\eta_{\text{SN}} =0.2$ is the mass fraction of stars going SNe and $m_*$
is the mass of the star particle.  We also return an amount $\eta_{SN}m_*$ back
into the gas for each SN explosion which occurs $10\,\rm Myr$ after the birth
of the star particle.  To avoid excessive cooling of the gas due to our
inability to capture the different phases of the SN bubble expansion, we use
the delayed cooling approach introduced in \citet{Teyssier+2013} (in the same
spirit as~\citealp{Stinson+2006}).  The energy of the SN explosion is injected
into a passive scalar variable and blocks the cooling of the gas if the
corresponding velocity dispersion is larger than  $\sigma _{\text{thres}} = 60
\, \rm km \, s^{-1}$.  The energy within that passive scalar decays with a
characteristic time-scale of $t_{\text{diss}}=2\,\rm Myr$
($t_{\text{diss}}=0.5\, \rm Myr$) for $\Delta x=40\,�\rm pc$ resolution
($\Delta x=10\,�\rm pc$ respectively), long enough to block the cooling over a
few cell sound crossing times~\citep[see Appendix of][]{duboisetal15}.
 
\subsection{Application of external pressure}
\label{sec:Pext}
 
\begin{figure}
\centering
\plotone{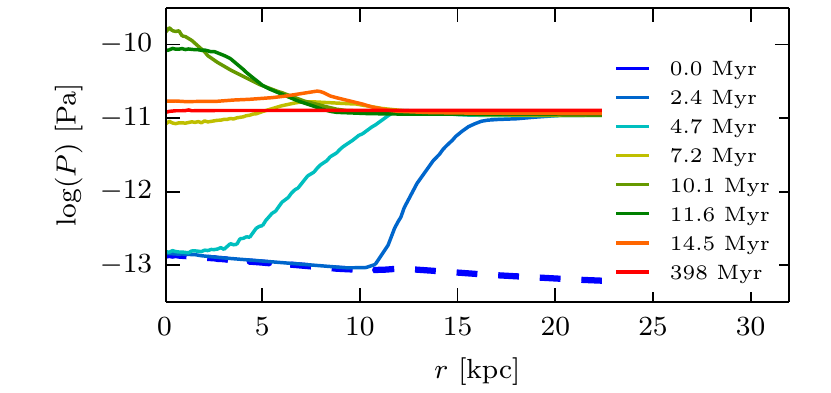}
\caption{Mean pressure versus radius at different times (see
legend) for the pa3 run of the \gasHighfb\ set.  The dashed line shows the
pressure profile before the onset of external pressure. The pressures are 
averaged within spherical shells. 
\label{fig:ProfileEvol}}
\end{figure}
 
After adiabatic relaxation (no gas cooling), the origin of time is reset to 0
and the base simulations are run further in time with the subgrid modeling of
gas cooling, star formation and feedback, and with an enhanced and uniform
pressure outside the disc (\pressure\ simulations) for another $\approx 420\,
\rm Myr$.  The pressure enhancement is applied at an instant starting at $t=0$.
This instant pressure increase is justified since  the bow shock observed in
the simulations of \citet{Gaibler+2012} manages to pressurise the entire
gaseous disc within a time-frame of only a few Myr.  The pressure is enhanced
in two different configurations: either outside the sphere of radius $r_1 =
12\,\rm kpc$ (hereafter, \pspher), or else where the gas number density is
lower than $0.014\, \rm H\, cm^{-3}$ (i.e. right outside the disc component,
hereafter, \pdens or equivalently \gasHighd). In the \pspher\ simulation, the
bow shock pressurising the disc is assumed to be quasi-isotropic. The effect of
isotropy of external pressure is compared with a simulation of a non-isotropic
bipolar pressure increase in Appendix~\ref{sec:Bipolar}.

In the case of spherical geometry (\pspher), this pressure enhancement is calculated by
\begin{equation}
P(r,t) = \left \{ \begin{array}{ll}
P(r,0) & r < r_1 \ , \\
\displaystyle {\rm pa} \,f\left({r-r_1\over r_2-r_1}\right)\,P_{\rm max} & r_1 \leq r < r_2 \ , \\
{\rm pa} \,P_{\rm max} & r \geq r_2 \ ,
\end{array}
\right.
\label{Pext}
\end{equation}
where time $t=0$ is just before the pressure enhancement, pa is the
\textit{pressure amplification}, $f(x) = 6 x^5-15 x^4+10x^3$ is an increasing
function of $x$ starting very gradually at $x=0$ and smoothly reaching a
plateau of unity at $x=1$, and finally $P_{\rm max}$ is the maximum pressure in
the disk at $t=0$ (reached in the central few cells), with $P_{\rm max} \simeq
9.8\times 10^{-13}\,\rm Pa$ for the \gasLow\ simulation set and  $P_{\rm max}
\simeq 4.7 \times 10^{-12}\,\rm Pa$ for the \gasHigh\ and \gasHighd\ simulation
sets.  Here, we adopt $r_2 = r_1 + 3 \,\rm kpc$.  This gradual pressure
amplification with radius is used to smoothly connect the two pressure regimes.
For convenience, we will call paX a simulation run where the pressure
amplification is ${\rm pa} = X$.  This pressure bath is maintained throughout
the simulation evolution and is a minimum to the pressure evolved in that
region. If the pressure within that bath becomes larger than ${\rm pa} P_{\rm
max}$ (due to SNe winds for instance) we take the new value of pressure
provided by the Riemann solver.  For this \pspher\ case (but also for
the \pdens\ case), the simulation of no pressure amplification corresponds to
${\rm pa} = P(r_1,0)/P_{\rm max} \simeq 0.1$, and we will hereafter denote it
as nP (for no pressure enhancement).

For the case of external pressure in disc geometry (\pdens), we increase the
pressure, only at time $t=0$, at a value of ${\rm pa} P_{\rm max}$ wherever the gas
density is below $0.014\, \rm H\, cm^{-3}$.  This gas density corresponds to a
height of 1.1 kpc above and below the plane along the minor axis of the disk
($R=0$).

In the simulations of \citet{Gaibler+2012}, the bow shock that pressurises the
disc reaches a maximum pressure of $P \simeq 8 \times 10^{-11}\,\rm Pa$.  This
justifies our chosen pressure enhancement where the maximum pressure increase
for the \gasHigh\ (pa10) and \gasLow\ (pa7) simulation corresponds to $P \simeq
9.8 \times 10^{-12}\, \rm Pa$ and $P \simeq 3.2 \times 10^{-11}\, \rm Pa$,
respectively.

The pressure profiles for one of the \pspher\ simulations (run pa3) before and
after the pressure enhancement as well as its evolution over the simulation
time are shown in Fig.~\ref{fig:ProfileEvol}.  We can see that that at $2.4\,
\rm Myr$, right after the restart of the simulation, the pressure smoothly
rises from $\sim 10^{-13} \, \rm Pa$ at the centre up to $10^{-11}\,\rm Pa$ at
a distance $r=13.5\, \rm kpc$.  At later times, this pressure enhancement
propagates within the central region of the halo and connects to the galaxy.
 
\begin{table}
\caption{Physical parameters of runs: gas fraction (0.1 for \gasLow\ and 0.5
for \gasHigh), pressure amplification (pa), run with no feedback (nf), run with
feedback (fb), spatial resolution ($\Delta x$), and pressure geometry } 
\begin{center}
\begin{tabular}{lrlcl}
\hline 
\hline
Identifier  & gas\ \ \  & \multicolumn{1}{c}{pa} & nf/fb & $\Delta x$ \qquad geometry \\ 
& fraction & & & [pc] \\
\hline
\multicolumn{2}{l}{pa01 $\equiv$ nP}  & 0.1 & \checkmark / \checkmark & 40  \\ 
pa04   &  & 0.4    & \checkmark / \checkmark & 40  \\
pa08    & & 0.8    & \checkmark / \checkmark & 40  \\
pa1.2 & & 1.2    & \checkmark / \checkmark & 40  \\
pa1.5  & \multirow{2}{*}{\qquad \gasLow} & 1.5    & \checkmark / \checkmark & 40 \qquad \multirow{2}{*}{\pspher} \\
pa3     & & 3      & \checkmark / \checkmark & 40  \\
pa5     & & 5      & \checkmark / \checkmark & 40  \\
pa7     & & 7      & \checkmark / \checkmark & 40  \\
\hline
\multicolumn{2}{l}{pa01 $\equiv$ nP}  & 0.1 & \checkmark / \checkmark & 40 \\
\multicolumn{2}{l}{pa01\_hR $\equiv$ nP\_hR}   & 0.1    & x / \checkmark & 10   \\
pa02   &  & 0.2 & \checkmark / \checkmark & 40  \\ 
pa04   &  & 0.4 & \checkmark / \checkmark & 40  \\ 
pa08   &  & 0.8 & \checkmark / \checkmark & 40  \\ 
pa1.2  &  & 1.2    & \checkmark / \checkmark & 40  \\ 
pa1.5  & \multirow{2}{*}{\qquad \gasHigh} & 1.5    & \checkmark / \checkmark & 40 \qquad \multirow{2}{*}{\pspher} \\
pa2     & & 2     & \checkmark / \checkmark & 40  \\ 
pa3     & & 3     & \checkmark / \checkmark & 40  \\ 
pa3\_hR & & 3     & x / \checkmark & 10   \\
pa5     & & 5     & \checkmark / \checkmark & 40  \\ 
pa7     & & 7     & \checkmark / \checkmark & 40  \\
pa8     & & 7     & \checkmark / \checkmark & 40  \\
pa10    & & 10    & \checkmark / \checkmark & 40  \\ 
\hline
\multicolumn{2}{l}{pa01\_d $\equiv$ nP}  & 0.1 & \checkmark / \checkmark & 40 \\
pa02\_d  &   & 0.2 & \checkmark / \checkmark & 40  \\ 
pa03\_d  &   & 0.3 & \checkmark / \checkmark & 40  \\ 
pa04\_d  &   & 0.4 & \checkmark / \checkmark & 40  \\ 
pa08\_d  &   & 0.8 & \checkmark / \checkmark & 40  \\ 
pa1.2\_d &  & 1.2    & \checkmark / \checkmark & 40  \\ 
pa1.5\_d  & \multirow{2}{*}{\quad \ \gasHigh} & 1.5    & \checkmark / \checkmark & 40 \qquad \multirow{2}{*}{\pdens}  \\
pa2\_d    & & 2      & \checkmark / \checkmark & 40  \\ 
pa3\_d    & & 3      & \checkmark / \checkmark & 40  \\ 
pa5\_d    & & 5      & \checkmark / \checkmark & 40  \\ 
pa7\_d    & & 7      & \checkmark / \checkmark & 40  \\ 
pa10\_d    & & 10      & \checkmark / \checkmark & 40  \\ 
\hline
\end{tabular} 
\end{center}
\label{tab:init2} 
\end{table}
 
\begin{figure*}
\plotoneF{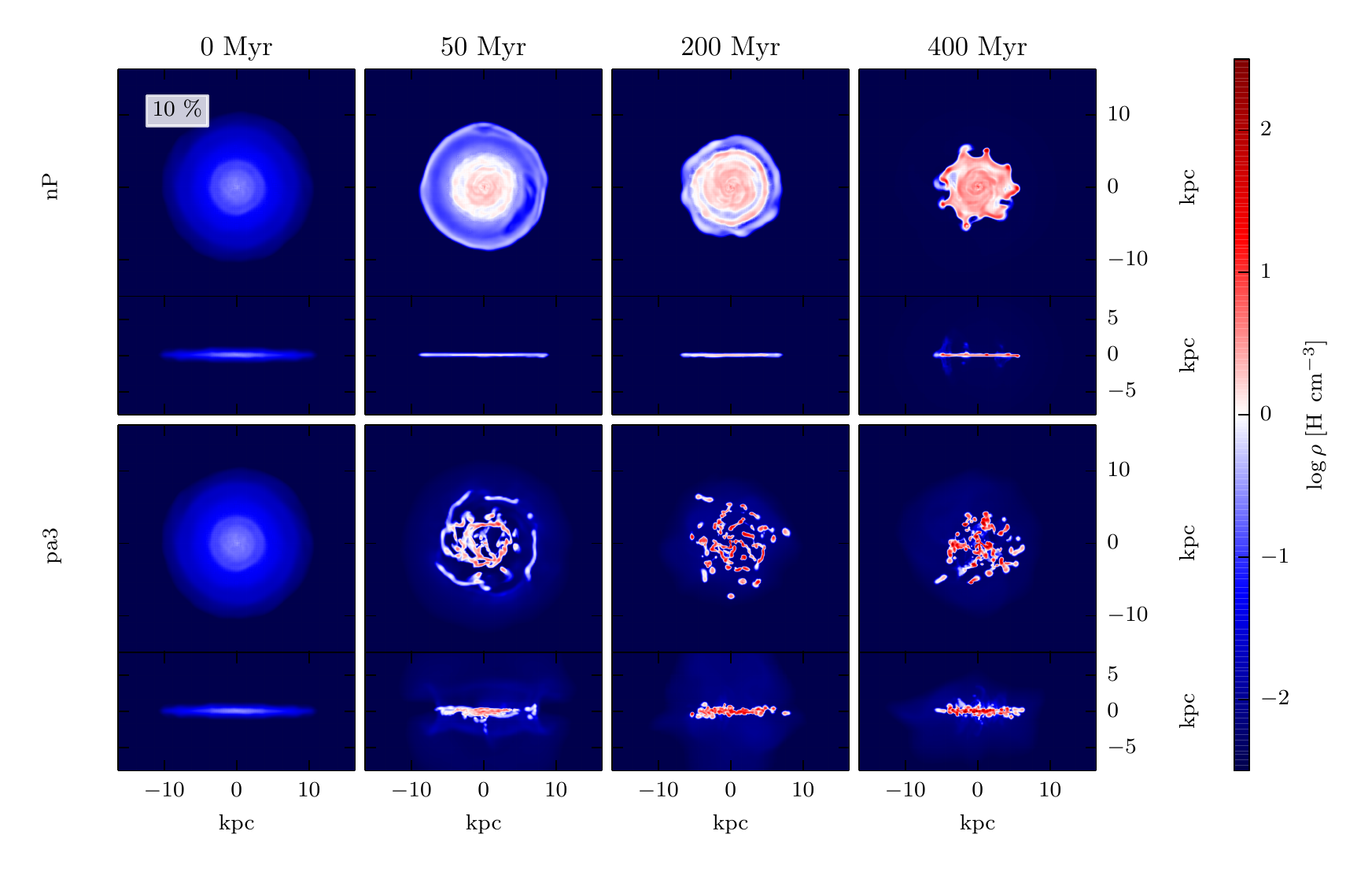}
\caption{Gas density maps (mass-weighted) of two of the \gasLowfb\ simulations
without enhancement of the external pressure nP (top row), and with enhancement
of the external pressure pa3 (bottom row).  The different columns show
different times as labelled. Each panel shows both face-on (40$\times$40 kpc,
upper part) and edge-on (40$\times$20 kpc, lower part) views. One can see that
an increased pressure outside the galaxy leads to accelerated clump formation
and less gas between the clumps.  }
\label{fig:Map_bulge1_ad_fb}
\end{figure*}
 
\begin{figure*}
\plotoneF{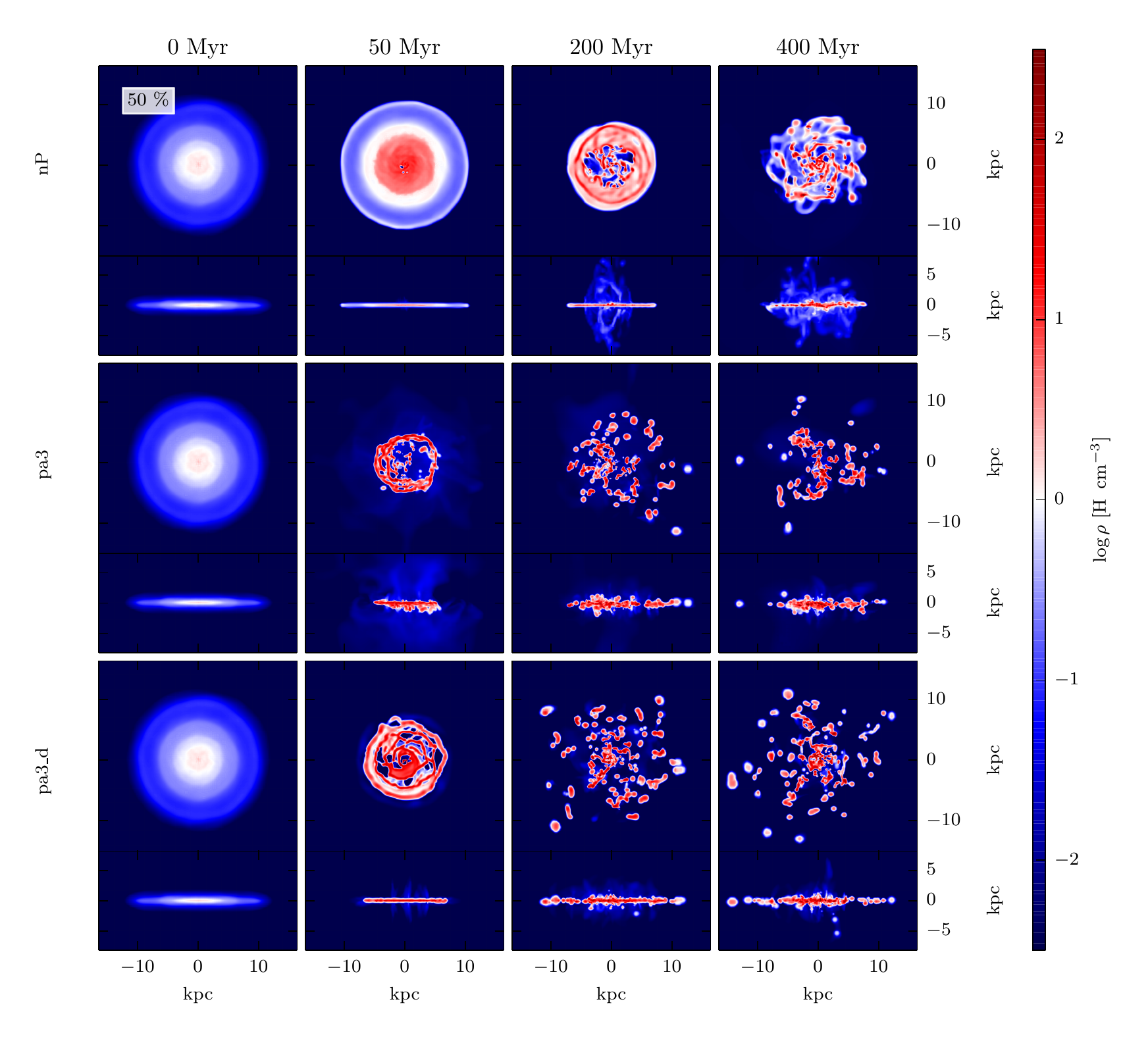}
\caption{Gas density map  (mass-weighted) for a selection of the \gasHighfb\
and \gasHighdfb\ simulations, for no pressure enhancement (top), and for a
pressure enhancement of pa3 (middle for \gasHighfb\ and bottom for
\gasHighdfb).  The density scale is as in Figs.~\ref{fig:Map_bulge1_ad_fb}.
An increased pressure outside the galaxy leads to accelerated clump formation
and less gas between the clumps. The morphological structure of the two
simulations with two different ways to increase the pressure (\gasHighfb, and
\gasHighdfb) is slighly different, but only in the outskirts of the galaxy.
The edge-on view shows a mass outflow for all the simulations. 
 }
\label{fig:Map_comp}
\end{figure*}
 
The relevant physical parameters for the pressure simulations are summarised in
Table~\ref{tab:init2}. 

\section{Results}
\label{sec:results}

In this section, we present our simulation results considering different
isolated disc simulations with various pressure boosts outside the disc. We
analyse our simulations regarding disc fragmentation, star formation, clump
properties, and the galaxy's mass budget.  We then compare our simulations with
a simple theoretical implementation regarding the growth of the star formation
rate and show that it scales approximately as the square root of the external pressure. 
Finally, we calculate the Kennicutt-Schmidt (KS) relation and find that our toy model for
AGN-induced over-pressurisation leads to the galaxies lying higher in the
starburst region of the KS relation. 

The  effects of external pressure turn out to be similar whether or
not SN feedback is included in the simulations.  We will therefore only
present, in this section, the results of
the stellar feedback simulations.  A comparison between the non-feedback and
feedback simulations is provided in Appendix~\ref{sec:SNnSN}.  

\subsection{Qualitative differences}
 
Figs.~\ref{fig:Map_bulge1_ad_fb} and \ref{fig:Map_comp} show maps of the
gas density for selected runs at different times, for the \gasLowfb\  and
\gasHighfb\ as well as \gasHighdfb\ runs respectively, and for two cases without
external pressure boost nP and with extra pressure pa3.  Comparing the nP
runs, we observe that the gas is clumpier in the \gasHighfb\ simulation than in
the \gasLowfb\ run. We will show in Sect.~\ref{sec:frag} that this is a simple
consequence of the Toomre instability.
The increased pressure leads to accelerated clump formation for the \gasLowfb,
\gasHighfb, \gasHighdfb\ simulations, and a clumpier ISM in all cases.
Generally less gas between clumps in the enhanced pressure runs,  in all the
\gasLowfb, \gasHighfb, and \gasHighdfb\ cases can be seen.  

In Fig.~\ref{fig:Map_comp}, one can compare the two different ways to increase
the pressure (\pspher, \pdens).  The morphological structure of the two
simulations \gasHighfb\ and \gasHighdfb\ is slightly different.  Fewer clumps
are seen in the \gasHighfb\ run than in the \gasHighdfb\ run. It seems,
however, that the clumps are only missing in the outskirts of the \gasHighfb\
galaxy, whereas  a similar amount of clumps can be detected in the centre.  The
edge-on views indicate that in the \gasHighfb\ simulations, a large amount of
mass flows out of the galaxy due to the pressure increase, while in the
\gasHighdfb\ simulations the mass outflow seems to be less extended.  We will
quantify the mass outflows in the different runs in
Sect.~\ref{subsec:MassFlow}. 

\subsection{Disc fragmentation}
\label{sec:frag}
 
\begin{figure*}
\plotoneF{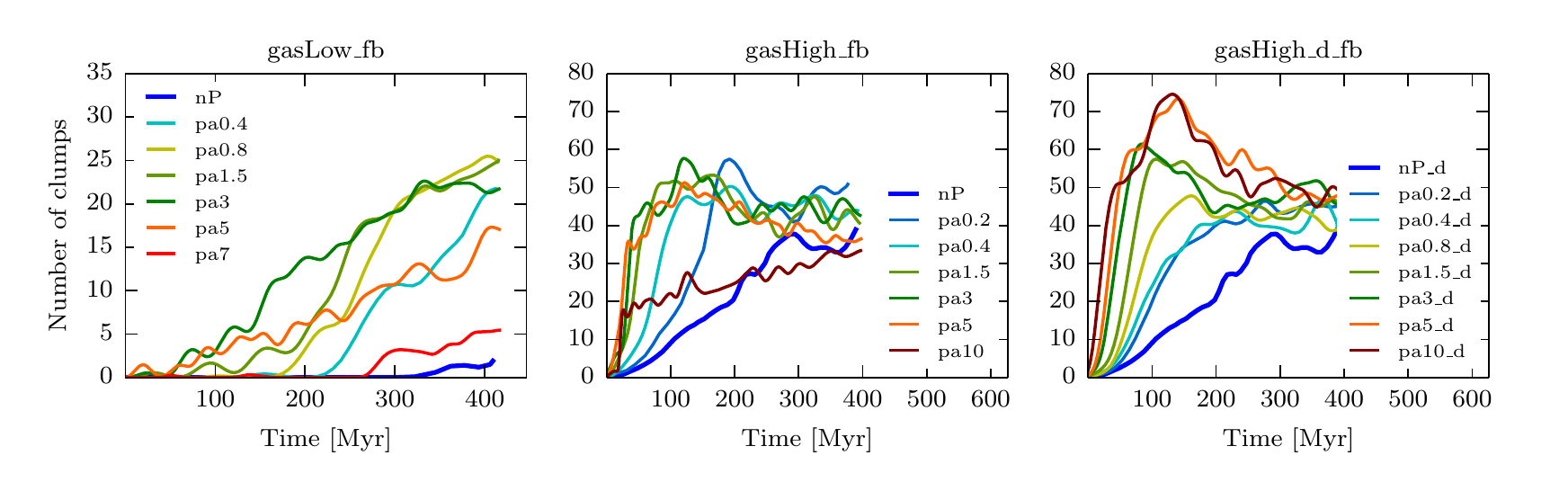}
\caption{Time evolution of the number of clumps for a selection of simulations
with supernova feedback: \gasLowfb\ (left), \gasHighfb\ (middle), and
\gasHighdfb\ (right) simulations.  The lines are smoothed with a
Blackman-Harris window with a width of $2 \sqrt{{\rm len(array)}}$, where
len(array) is the number of points.  The clumps were extracted with the
\citet{Bleuler+Teyssier2014} algorithm, with a density threshold of 21~$H\, cm^{-3}$
and a peak-to-saddle threshold of 1.5. The maps show that an increased
pressure leads to increased clump formation and the increase in clump number is
dependent on the pressure applied onto the galaxy. Beyond a certain pressure
enhancement, the number of clumps decreases or remains very similar for higher
pressure runs. } 
\label{fig:Number_Clump}
\end{figure*}
\begin{figure}
\plotone{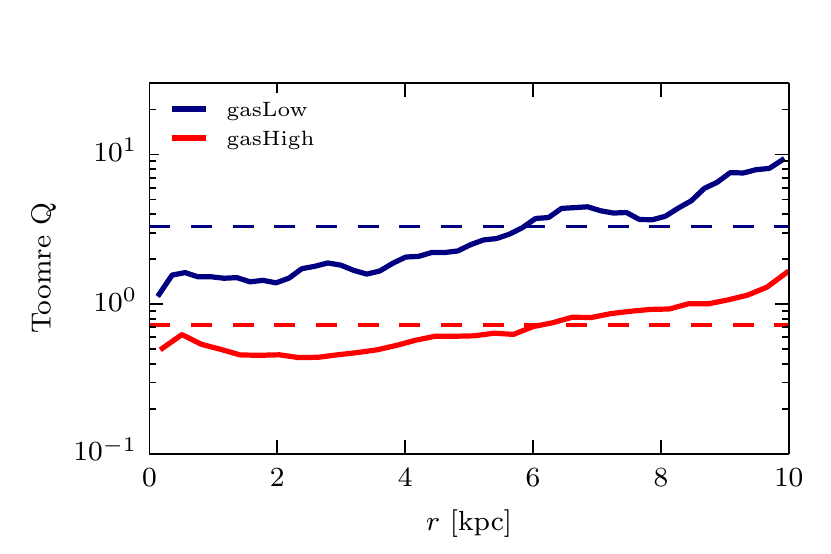}
\caption{Local Toomre parameter for the relaxed disc at $t=0$ for the \gasLow\
and \gasHigh\ simulations. The dashed line shows the mean Toomre parameter of
the disc  for the \gasLow\ simulation  $\langle Q \rangle =3.29>1$ and
for the \gasHigh\ simulation $\langle Q\rangle =0.72 < 1$. One therefore expects
the \gasHigh\ simulation to fragment independently of external pressure
enhancement whereas the \gasLow\ simulations are not expected to
fragment into many clumps. The fact that  the pressur-enhanced
simulations of the \gasLow\ galaxies shows a significant increase in
the number of clumps compared to no-pressure enhancement demonstrates that
external pressure can stimulate the fragmentation of a disc even if it is 
Toomre-stable.} 
\label{fig:Toomre}
\end{figure}
 
Since the star formation recipe depends on the local gas density (see
eq.~\ref{eq:sfr}), we expect enhanced star formation when more clumps are
formed (as the gas gets more concentrated), assuming that the clumps have
sufficient mass. Therefore, if an increased pressure leads to increased
fragmentation and hence increased clump formation, we expect star formation
to be positively enhanced when external pressure is applied to the galaxy.  We
first consider the fragmentation by counting the high-density clumps.  We
detect the clumps in the simulation by running the clump finder described by
\citet{Bleuler+Teyssier2014}. This method identifies all peaks and their
highest saddle points above a given threshold ($21\,\rm H\, cm^{-3}$). A clump
is recognised as an individual entity when the peak-to-saddle ratio is greater
than 1.5; otherwise the density peak is merged with the neighbor peak with
which it shares the highest saddle point.  

The visual impression of increased clump formation when external pressure is
applied on the galaxy (Figs.~\ref{fig:Map_bulge1_ad_fb} and
\ref{fig:Map_comp}), is confirmed when looking at the number of clumps as a
function of time.  Fig.~\ref{fig:Number_Clump} shows the number of clumps as a
function of time for the \gasLowfb\ (left panel), \gasHighfb\ (middle panel),
and \gasHighdfb\ (right panel) simulations, respectively. 

In the \gasLowfb\ run, the number of clumps is constantly increasing with time,
regardless of the amount of external pressure.  Clump formation starts earlier
in the runs with external pressure.  However, the number of clumps at a given
time is not a monotonic function of external pressure: at low external pressure
(up to pa3), the number of clumps at given time increases with increasing
pressure, while the reverse trend occurs for external pressures above $3\,P_{\rm
max}$.

The general effect that more clumps are formed in the simulations with external
pressure is similar for the \gasHighfb\ and \gasHighdfb\ runs. Similar to the
\gasLowfb\ simulation, the number of clumps increases with increasing pressure
up to a certain pressure (pa5) and then decreases again for the \gasHighfb\
simulation and stays at the same level for the \gasHighdfb\ simulation.  For
the lower pressure as well as the non-pressure simulations, the number of
clumps increases with time.  However, for higher pressure simulations, the
number of clumps reaches a plateau at late times.  For the \gasHighfb\ runs,
the initial rise in the number of clumps is fastest for the pa3, pa5 and pa7
cases, but in the pa3 case the number of clumps reaches its plateau at a later
time, hence at a higher level.

The time evolution of the number of clumps for the \gasHighdfb\ simulation is
roughly similar to the \gasHighfb\ simulation.  At early times, the rise in
number of clumps is fastest for the high pressure runs.  The number of clumps
keep rising with time for the lower pressure enhancements, while  it reaches a
maximum for the higher pressure enhancements. The time when the number of
clumps reaches its plateau is also shortest for higher external pressures.
After 300~Myr, there is no clear trend in number of clumps versus external
pressure for the higher pressure runs.  The increase of clump number is
therefore highly dependent on the pressure applied on the galaxy. However,
beyond a certain pressure enhancement (pa5\_d), the number of clumps remains
very similar for higher pressure runs.  After $\approx$~300~Myr the number of
clumps is roughly independent of external pressure for the \gasHighfb\ and
\gasHighdfb\ simulations.

In the \gasLowfb\ run with no external pressure, only a single clump in the
entire disc is formed, at late times (350~Myr), when the gas has sufficiently
collapsed to reach the clump gas density threshold.  On the contrary, in the
\gasHighfb\ run, the number of clumps increases up to $\simeq 35$, even without
any forcing by the external pressure.  

The difference between the gas-poor and gas-rich galaxies, before the external
pressure is applied, is that the gaseous disc is Toomre-stable against
small-scale fragmentation in the gas-poor case where the mean Toomre parameter
is $\langle Q \rangle =\langle c_{\rm s}\,\kappa/(\pi G \,\Sigma_{\rm gas})
\rangle = 3.29>1$, while the gas-rich disc is Toomre-unstable with $\langle
Q\rangle =0.72 < 1$ (see Fig.~\ref{fig:Toomre}).  Here, $\Sigma_{\rm gas}$ is
the surface density, $c_{\rm s}$ is the sound speed, and $\kappa$ is the
epicyclic frequency (measuring the shear of the rotating disc).  Therefore the
\gasLowfb\ simulations demonstrate that fragmentation of the galactic disc can
be driven by the forcing of an external pressure, even though the disc is
initially Toomre-stable. 

\subsection{Star formation history}

\begin{figure*}
\plotoneF{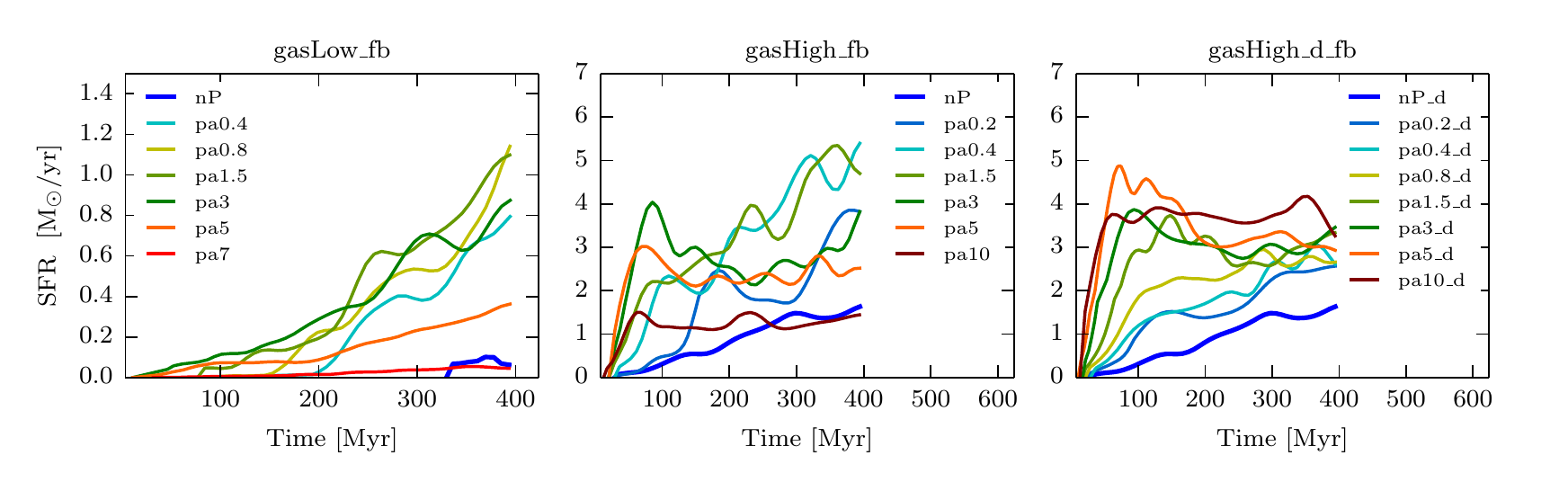}
\caption{SFR for a selection of the simulations with supernova feedback:
\gasLowfb\ (left), \gasHighfb\ (middle), and \gasHighdfb\ (right) simulations.
The lines are smoothed as in Figure~\ref{fig:Number_Clump}. This figure shows
that higher pressures lead to higher SFRs.  For the higher pressure enhancement
simulations, the SFR reaches a maximum or plateau at later times for the
\gasHighfb, and \gasHighdfb\ simulations. After a certain pressure increase, the
SFR decreases or stays at the same level for all the simulations and hence the
runs with intermediate pressure  generally produce the highest SFR at all times.
}
\label{fig:SFR}
\end{figure*}
 
In Sect.~\ref{sec:frag}, we have seen that an increased pressure enhancement
leads to an increased number of clumps up to a certain time and then a
typically lower number of clumps thereafter.  Since the gas density threshold
of clump detection is set to be above that for star formation, one expects that
the star formation history should evolve in a similar fashion to the evolution
of the number of clumps. 

Fig.~\ref{fig:SFR} shows that the star formation histories of the different
runs indeed resemble the time evolution of the number of clumps previously
shown in Fig.~\ref{fig:Number_Clump}.  In particular, at early times in the
runs with \gasHigh\, higher pressures lead to higher SFRs.  But with
high pressures, the SFR saturates earlier. In the \gasHighfb\ runs, the maximum
SFRs in the high pressure runs are lower than in the other runs, while in the
\gasHighdfb\ runs, the maximum level of SFR is reached for the three highest
pressures, while the SFRs at later times (300~Myr) are roughly independent of
the external pressure.

In the \gasLowfb\ runs, while the nP case leads to star formation only after a
long time delay (330 Myr), the highest pressures, although leading to immediate
but small levels of star formation, are unable to generate substantial star
formation from the earliest times.  The runs with intermediate pressures
produce the highest SFR at all times. We will show in the next sections that
this is due to the low mass-outflow of the intermediate pressure simulations
that allows the clumps to increase in density.  Conversely, larger external
pressures lead to such strong pressure waves that the gas is removed from the
galaxy. This prevents the formation of large clumps and tends to suppress the
star formation. 
    
The effect of the external pressure on the SFR is even more significant when
looking at the \gasHighfb\ simulations (middle panel of Fig.~\ref{fig:SFR}).
The SFR of the no-pressure simulation slowly increases after a certain time,
whereas the SFR increases faster when pressure is applied: it reaches a maximum
at a certain rate and more or less maintains this rate for the remaining of the
simulation.  Towards the end of the simulation, the SFR of the no-pressure
simulation catches up, again similarly to the clump number behaviour.  The SFR
for the \gasHighdfb\ simulation (right panel of Fig.~\ref{fig:SFR}) behaves
quantitatively similar to the \gasHighfb\ simulation.  One can see that the
SFR in these simulations reaches the maximum or plateau at later times than
in the corresponding \gasHighfb\ simulations.  The rapid rise of the SFR
reaches increasingly higher levels of peak SFR with higher external
pressure up to pa5\_d, while pa10\_d reaches a slightly lower maximum
SFR.

The left panels of Fig.~\ref{fig:SFR} show that the SFRs of the \gasLow\
simulations start with a significant time delay, and the maximum enhancement of
the SFR relative to the nP run is highest ($\sim$~12) at the end of the
simulation (after $400$~Myr). On the other hand, the corresponding SFR
enhancements for the higher gas fraction simulations (middle and right panels
of Fig.~\ref{fig:SFR}) are lower ($\sim$~3.5 for \gasHigh\ and $\sim$~1.5 for
\gasHighd) at the end of the simulation (after $400$~Myr) than at the beginning
($\sim$~40 for \gasHigh\ and $\sim$ 70 for \gasHighd\ at $\sim 80$~Myr) of the
simulation.  External pressure thus first produces a significantly higher SFR
in comparison to the simulation with no external pressure.  But the duration of
this large SFR enhancement for the \gasHigh\ and \gasHighd\ simulations is
shorter than that of the \gasLow\ simulation. The free fall time of the higher
density gas is shorter than the free fall time of the low density gas which
leads to the gas collapsing early on, whereas a delay is expected for the lower
gas fraction disc.  

\subsection{Clump Properties}
\label{sec:clumpProps}
 
\begin{figure*}
\plotoneF{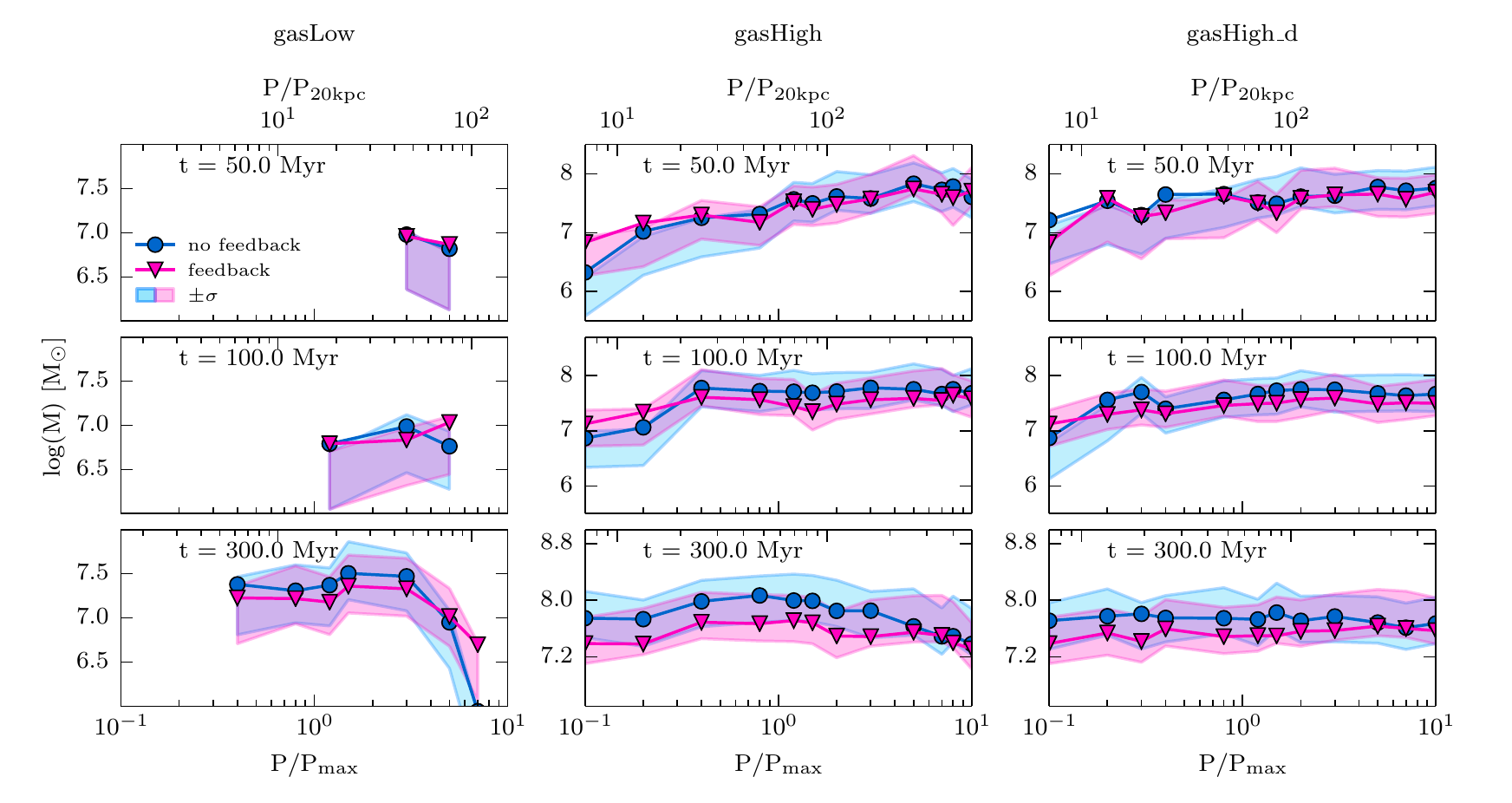}
\caption{Average clump mass for the \gasLowfb\ (left), \gasHighfb\ (middle),
and \gasHighdfb\ (right) simulations. The orange line with round markers
correspond to the non-feedback simulations whereas the pink line with triangles
corresponds to the feedback simulations. The shaded green/blue area shows the
area of mass containing $\pm \sigma$ of the density PDF for the
no-feedback/feedback simulations.  In the bottom and top of each subfigure, the
x-axis shows the $P/P_{\rm max}$ and $P/P_{\rm 20\, kpc}$ values
respectively, where $P_{\rm max}$ is the maximum pressure inside the disk
and $P_{\rm 20\,kpc}$ is the averaged pressure at 20~kpc. At the
beginning of the simulation, the clump masses for the \gasHigh\
simulations are higher the greater the pressure for the \gasHigh\
simulations. At later times, the clump masses are roughly independent of the
external pressure. For the \gasLow\ simulation, the clump mass does not
increase with higher external pressure but rather decreases or stays at
approximately the same level.} 
\label{fig:ClumpPropMass}
\end{figure*}
 
An important SFR requires a significant supply of cold gas as well as the
fragmentation of the disc into clumps that carry a sufficient amount of gas to
form stars. On the other hand, one can argue with the Jeans and Toomre instability
arguments if indeed an increased pressure outside the galaxy that later increases
the pressure inside the galaxy leads to higher densities, as well as a possible
expulsion of disc gas depending on the momentum carried by the pressure wave
coming into the disc.  The competition between higher densities and mass
outflow will influence the amount of gas within the clumps. 

For the gas-rich disc simulations, we saw (Fig.~\ref{fig:Number_Clump}) that,
at the very beginning, when the pressure wave comes into the galaxy, the number
of clumps is highest for the highest pressure. While the clumps are more
numerous with the highest pressures, it is worthwhile knowing whether their
masses are affected by the external pressure.

\begin{figure*}
\plotoneF{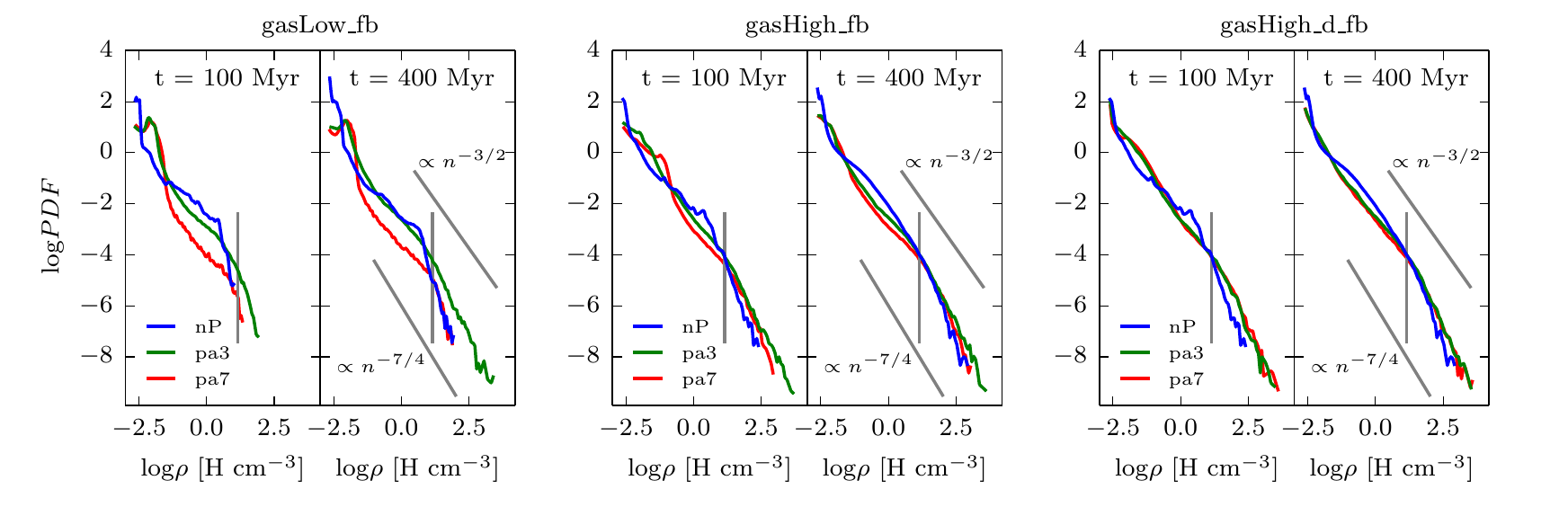}
\caption{Density PDF at different times for a selection of the \gasLowfb\
(left), \gasHighfb\ (middle) and \gasHighdfb\ (right) simulations.  The
threshold ($14\, \rm H\, cm^{-3}$) for star formation is plotted as a grey
vertical line.  One can see that greater external pressure allows the galaxy to
reach higher densities on a faster time-scale. The nP simulation slowly catches
up with the over-pressure simulations for the \gasHigh\ whereas this is not the
case for the \gasLow\ simulation. For the \gasLow\ simulation, the high
pressure simulation (pa7)  never reaches the high densities. As indicated with
the grey power law lines one can see that over the course of the simulation a
slope between -3/2 and -7/5 (or even steeper) develops at high densities.  This
is in agreement with simulations including gravity and turbulence done for
instance by \citet{Kritsuk+2011}.}
\label{fig:PDF}
\end{figure*} 

Fig.~\ref{fig:ClumpPropMass} shows the modulation of the average clump mass
rises with external pressure at three times of the simulations.  At early times
(top panel), the clump masses for the \gasHigh\ simulations (\gasHigh, center
panel, and \gasHighd, right panel) are higher the greater the external
pressure. At later times (middle and bottom panel), the clump masses are
roughly independent of the external pressure applied, probably because the disc
gas has been either already accreted onto the clumps or expelled out of the
galaxy (see discussion below), leaving no more diffuse gas available for
accretion onto the clumps. The time at which the diffuse gas is either consumed
onto the clumps or expelled must happen earlier for the higher external
pressure simulations as the fragmentation happened earlier for these
simulations. This explains the different times when the SFR reaches a plateau,
occurring earlier the higher the pressure.  In the \gasHighfb\ runs at high
external pressures, fragmentation is not the only cause of SFR (since there is
a maximum pressure enhancement beyond which the SFR is lower),  meaning that
the gas supply is more critical, and not always available despite the high gas
fraction. This suggests that the mass flow out of the galaxy also plays an
important role.  And indeed, the mass outflow is very efficient for pa7 and
pa10 after 30~Myr. We will discuss this in detail in
Sect.~\ref{subsec:MassFlow}, below.
 
In the \pdens\ simulations,  there is a maximum pressure enhancement (pa5\_d)
beyond which the SFR remains at approximately the same level without
decreasing.  As we will see in Sect.~\ref{subsec:MassFlow}, mass outflows are
also absent.  The high gas fraction leads therefore to higher density
enhancement by external pressure, hence both number of clumps and SFR are
highest when the external pressure are high. However, there only is a limited
amount of gas available in the galaxy.  One can assume that the limited gas
supply is insufficient for more star formation, so the SFR remains at the same
level independently of the pressure enhancement.

Fig.~\ref{fig:ClumpPropMass} shows that, for the \gasLow\ simulations (\gasLow,
left panel), the clump mass does not increase with greater external pressure,
but rather decreases at early times (top panel). At later times, the clump mass
is roughly independent of the pressure up to pa3, beyond which the clump mass
decreases.  This is the same pressure enhancement which leads to the highest
SFR. As we will see in Sect.~\ref{subsec:MassFlow}, the \gasLow\ galaxies
suffer from strong gas outflows that reduce the supply of gas available for
clump buildup, leading in turn to smaller clump masses within the galaxy.

In Fig.~\ref{fig:ClumpPropMass}, the difference in clump masses for the
feedback and non-feedback simulations can also be seen for all the simulations.
At early times, there is a significant difference between the no-feedback and
feedback simulations. At later times, the difference in clump masses becomes
more apparent for both the high gas fraction and \gasLow\ simulations.  One can
see that the clump masses for the feedback simulations are lower at the end of
the simulation than for the no-feedback simulations, independent of the
pressure increase. Because the feedback increases the porosity of the
interstellar medium that in turn counteracts the formation of clumps
\citep{Silk2001}, the observed smaller clump masses for the feedback
simulations are expected.  This difference is more dominant in the \gasHigh\
simulation as can be seen in the left panel of Fig.~\ref{fig:ClumpPropMass}.
The similar mean clump masses in the feedback and non-feedback runs at the
beginning of the simulation appears to be a consequence of the implementation
of the SNe in the simulation.  As discussed above, a SN explosion occurs
$10$~Myr after the birth of the star particle. The first stars form shortly
before $50$~Myr and one would therefore not expect to see a large difference
between the feedback and no-feedback simulations. At $100$~Myr, some stars
exploded into SNe, but only constitute a small fraction of all stars, hence the
small difference between the feedback and no-feedback simulations at this
stage.

It is interesting to look at the density probability function (PDF) at
different times of the simulations in order to better understand the observed
SFR behaviour.  The PDF can be seen in Fig.~\ref{fig:PDF} for two different
times for a selection of the \gasLow\ (left), \gasHigh\ (middle), and
\gasHighd\ (right) simulations. Increasing external pressure allows one to
reach higher gas densities faster, which is in agreement with the SFR behaviour
we have seen previously. For the \gasHigh\ simulations, the nP simulation
slowly catches up with the over-pressure simulations similar to the SFR
behaviour. For the \gasLow\ simulation, the no-pressure simulations never
attain the densities reached by   the moderate pressure enhancement
simulations. The high pressure simulation (pa7) also never reaches high gas
discuss in Sect.~\ref{subsec:MassFlow}, below.  It can be seen in
Fig.~\ref{fig:PDF} that over the course of the simulation a high density power
law with a slope between -7/4 (or even steeper) and -3/2 develops, especially
for the high gas fraction simulations. A comparable power law range has been
found in observations (e.g., \citealp{Kainulainen+2009},
\citealp{Lombardi+2010}) and simulations including gravity and turbulence
(e.g, \citealp{Kritsuk+2011}) where they argue that the origin of the power
law tail is due to self-similar collapse solutions. 

\subsection{The Galaxy's Mass budget}
\label{subsec:MassFlow}
 
\begin{figure*}
\plotoneF{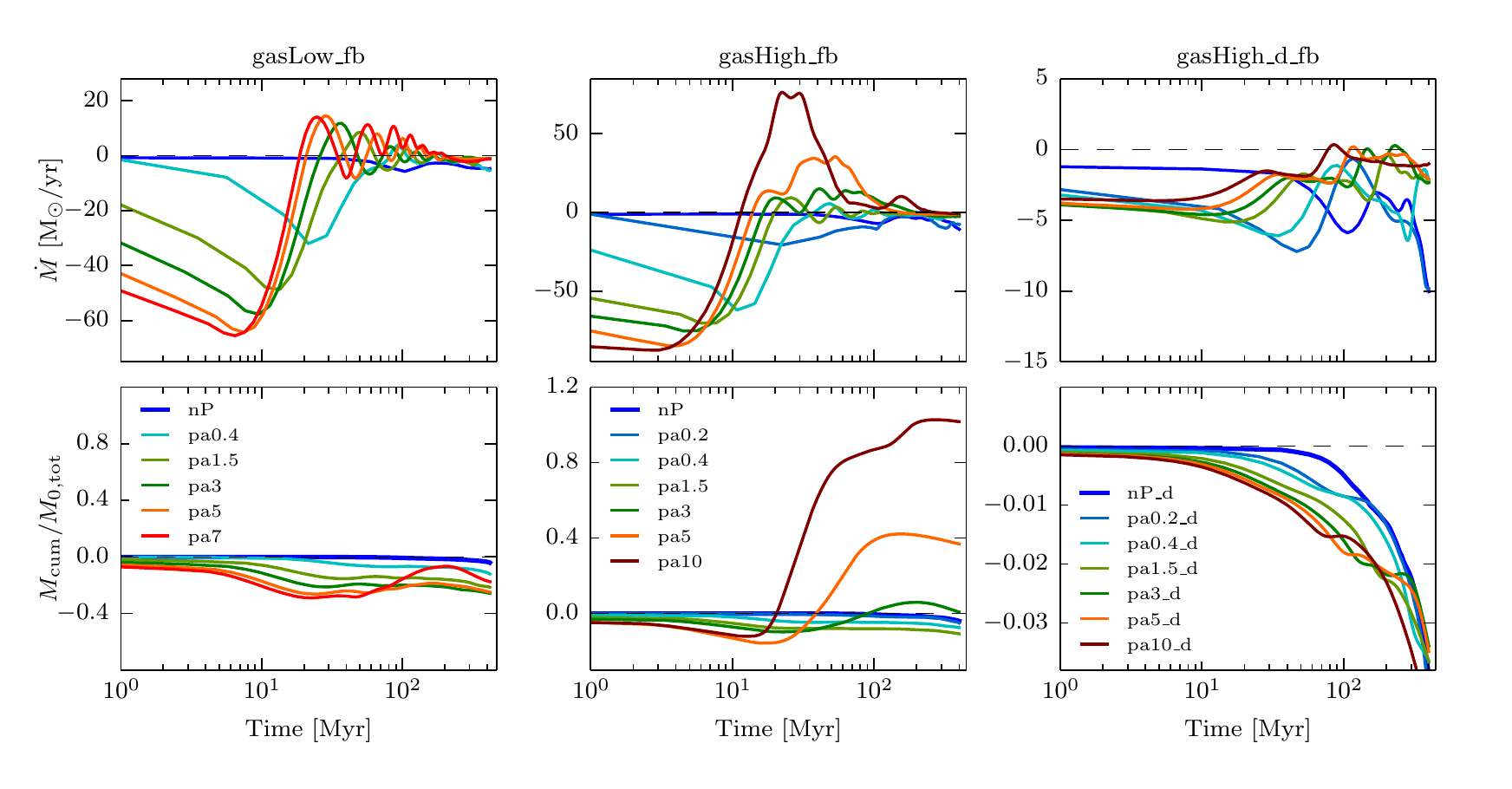}
\caption{Time evolution of the mass flow rate (top) and cumulative mass flow
relative to the initial mass $M_{\rm 0, tot}$ before inputting the pressure
(bottom), both at 16~kpc from the galaxy center (where the initial mass is
within a sphere of that radius), for selected runs from the \gasLowfb\
(left), \gasHighfb\ (middle) and \gasHighdfb\ (right) simulations.  Negative
(positive) values of the mass outflow rate denote a net mass inflow (outflow).
One can see a difference in the mass flow rate (MFR) between the different ways
pressure is put on the galaxy (\gasHighfb, \gasHighdfb).  Due to the pressure
gradient in the \pspher\ simulations, the pressure wave coming into the galaxy
carries a lot of momentum that leads to a mass inflow followed by a mass outflow
for the higher pressure simulations. This mass outflow is negligible for
the \pdens\ simulations due to the pressure wave carrying very little momentum.
For the most extreme case, the expelled mass reaches 80\% of the initial gas
mass for the highest external pressure simulation of \gasHighfb.}
\label{fig:MassOutflow}
\end{figure*} 
 
The mass flow rate (MFR) as well as the total amount of newly formed stars plus
dense gas should provide us a better understanding of the star formation
history described above.  In particular, one would like to understand why there
seems to be an optimal external pressure enhancement for star formation,
beyond 
which the SFR ends up at lower values.
 
We measure the gas mass flux through a sphere of radius 16~kpc as
\begin{equation} 
\dot{M}_{\rm gas} = \oiint \rho \,\vec{v} \cdot \hat{\vec{r}}\, \mathrm{d}S 
= \sum_{i \in \mathrm{shell}} m_{i}\, \vec{v}_{i} \cdot \hat{\vec{r}}_i
\,{S\over V}
\ ,
\label{massflux}
\end{equation} 
where $i$ denotes the index of a cell within a spherical shell of surface $S$
and volume $V$. Here, we adopt a shell of thickness 4 kpc.  The MFR is shown in
Fig.~\ref{fig:MassOutflow} for the \gasLowfb\ (left), \gasHighfb\ (middle), and
\gasHighdfb\ (right) simulations, again only for the simulations with SN
feedback.  
The top panels show the mass flow, while the
bottom ones show the cumulative mass flow in fractions of the total gas mass
within the 16~kpc  sphere before the pressure increase shown on the bottom. 
We will discuss the effects of SN feedback in Appendix~\ref{sec:SNnSN}.

In all three sets of simulations, external pressure leads to mass inflow at
early times. This early mass inflow is large but different for the different ways
pressure is applied onto the galaxy. In the \pspher\ simulations, the pressure
is applied outside the galaxy in a low density medium leading the pressure to
have a larger pressure gradient than in the \pdens\ simulations where the
pressure is applied close to the galaxy and therefore in a higher density
environment. This larger pressure gradient in the \pspher\ simulations,
allows the pressure wave to carry more mass and  momentum from the ambient hot
medium in comparison to the pressure wave of the \pdens\ simulations. This explains the 
larger mass inflow observed in the \pspher\ simulations (left and middle panels of
Fig.~\ref{fig:MassOutflow}) compared to that at the start of the
\pdens\ simulation. With its larger momentum, the mass inflow of the
\pspher\ pressure wave is
followed by a short period of mass outflow (for both low and high gas
fractions).
This mass outflow is negligible
for the \gasHighdfb\ simulations, since the pressure wave carries very little
momentum. 

For the \pspher\ simulation sets, higher external pressures lead to stronger
maximum inflows at early times and to stronger maximum outflows at later times.
In addition, in the simulations with high external pressures (pa7 and pa10), the
mass outflow that follows the mass inflow occurs very rapidly (in less than 20
Myr). After these two phases of important mass inflow/outflow, the MFR
oscillates around zero for both the \gasLowfb\ and \gasHighfb\ simulations 
with $\rm pa < 5$. In contrast, in the \gasHighdfb\ simulations,
the MFR depends little on the external pressure.

\begin{figure*}
\plotoneF{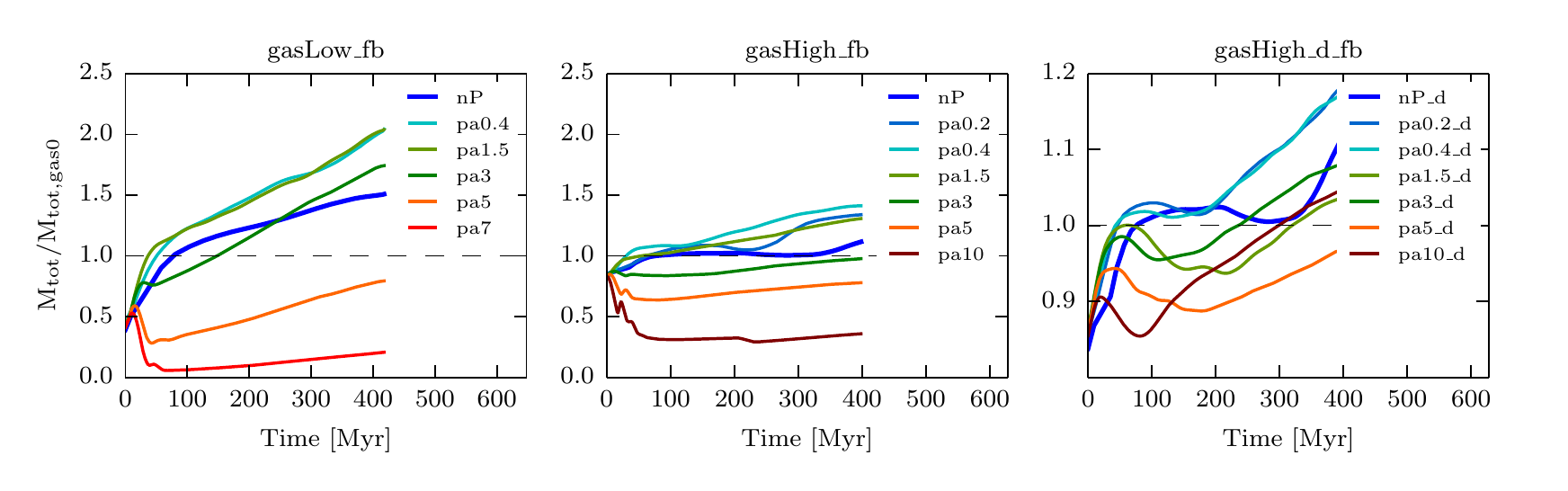}
\caption{Time evolution of the mass in newly formed stars plus dense ($n >
0.1\,\rm H\, cm^{-3}$) relative to the initial gas mass for a selection of the
simulations with SN feedback: \gasLowfb\ (left), \gasHighfb\ (middle), and
\gasHighdfb\ (right) simulations.  The lines are smoothed as in
Figure~\ref{fig:Number_Clump}.  Due to the mass inflow shown in
Fig.~\ref{fig:MassOutflow}, at the beginning of the simulation more mass can
end up in the galaxy. This extra mass of gas is more significant for the
\gasLow\ galaxy for both the no-pressure and low-pressure enhancement
simulations. For high pressure enhancement in the \gasLow\ simulation, the
incoming pressure wave significantly disperses the galactic gas. The evolution
of $M_{\rm tot}$ is similar for the \gasHighfb\ simulation albeit with less
mass variation. For the \gasHighdfb\ simulations, no significant mass variation
within the galaxy due to pressurisation is observed.}
\label{fig:totalMass}
\end{figure*} %

It is also instructive to consider the evolution of the cumulative gas mass
flow through the 16~kpc sphere (bottom panels of Fig.~\ref{fig:MassOutflow})
shown relative to the initial mass within the sphere of that radius.  The
cumulative mass flow remains negative (e.g. mass inflow) for all \gasLowfb\
simulations.  The cumulative mass flow for the \gasHighfb\ with low external
pressures (pa0.2 and pa0.8) remain negative, while  for higher external
pressures, they end up positive.  Whereas for low external  pressure ($\leq$
pa3) the cumulative mass outflow is less than 5\% of the initial gas mass, the
cumulative mass outflow reaches 80\% of the initial gas mass for the highest
external pressure simulation pa10.  Finally, all simulations in the \pdens\
geometry lead to cumulative mass inflow at all times, with the strongest
cumulative inflows occurring for the runs with the greatest external pressures.
We  stress, however, that the pressure and 
no-pressure \pdens\ simulations do not differ significantly and that, overall,
there is little net mass flow.  This most likely explains why the SFR of the
\pdens\ simulations is smoother and less noisy when compared to the \pspher\
simulations of the same gas fraction (\gasHigh).  

In order to understand the galaxy's mass budget better we look at the time
evolution of the total mass of newly formed stars plus dense ($n>0.1\,\rm H\,
cm^{-3}$) gas, $M_{\rm tot} = M_{\rm tot,starsN} + M_{\rm tot,gasD}$ is shown
relative to the initial galaxy gas mass (total not just dense), $M_{\rm
tot,gas0}$ (see Table~\ref{tab:init1}) shown in Fig.~\ref{fig:totalMass}.  The
initial value of $M_{\rm tot}/M_{\rm tot,gas 0}$ is below unity at $t=0$ since
the gas density in the galaxy is not everywhere above $n>0.1\,\rm H\, cm^{-3}$,
especially in the outskirts of the disc and for the \gasLow\ galaxy.  In the
absence of extra external pressure (nP runs), the ratio $M_{\rm tot}/M_{\rm
tot,gas 0}$ quickly moves significantly beyond unity as the gas cooling allows
to reach the gas densities above $n>0.1\,\rm H\, cm^{-3}$.  The gas cooling
also takes place in the circumgalactic medium that feeds the galaxy with some
extra gas.  This extra mass of gas adds more significantly to the low-gas
fraction galaxy because of its lower initial gas mass, which explains why the
increase is more significant in the \gasLow\ runs than in the \gasHigh\ runs.

For the \gasLow\ simulations, intermediate regimes of forced external
pressure (from pa0.4 to pa3) also show values of $M_{\rm tot}/M_{\rm tot,gas 0}$
above 1 with values comparable to the nP run.  Therefore, the increase in SFR
(an order of magnitude above nP) is to be attributed to the extra compression
of the ISM and exploration of larger gas densities with shorter collapsing 
time-scales (see Fig.~\ref{fig:PDF}).  In contrast, higher pressurisation values
of the ISM (pa5 and pa7) lead to strong gas removal due to the incoming
pressure wave that manages to significantly disperse the galactic gas.  Since
the gas reservoir is reduced, the SFR is also suppressed compared to more
intermediate regimes of pressurisation, but the overall SFR is still larger
than in the nP case, where gas fragmentation is not reached.

The evolution of $M_{\rm tot}/M_{\rm tot,gas0}$ in the \pspher\ \gasHigh\ galaxy
behaves similarly to that for the \gasLow\ galaxy, although with lower
mass variation.  It starts below unity for all pressures, and decreases even
more  for the high pressure increases (pa5 and
pa10), because of the large mass outflows observed for those runs. This shows
that the large mass outflows associated with high pressures  prevent star
formation.  
On the other hand, $M_{\rm tot}/M_{\rm tot,gas0}$
keeps rising for the lower pressure enhancements, showing that because no large
mass outflow is observed, more stars can be formed.  The $M_{\rm tot}/M_{\rm tot,gas0}$
curve is higher when a small pressure is applied outside the galaxy compared with
the no-pressure simulation.  In the case of \pdens\ over-pressurisation, there
is no significant ($<20$ percent relative) mass variation in the galaxy.  
Thus, the early fragmentation due to the increased
pressure drives the different SFR levels for the different pressure simulations.  

\subsection{The Star Formation Rate}
\label{sec:dSFR}
 
\begin{figure*}
\plotoneF{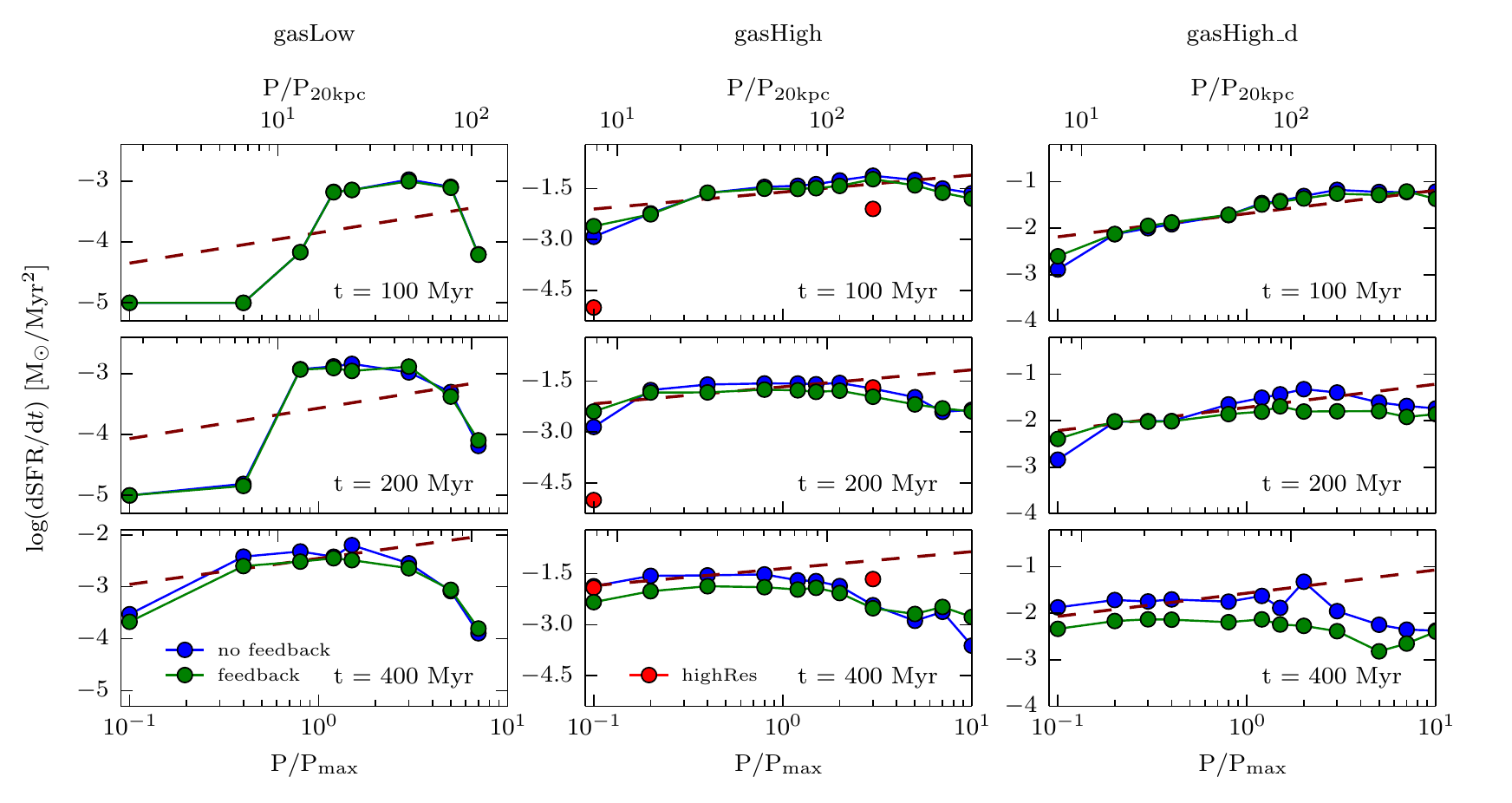}
\caption{Time evolution of the average rate of increase of the star formation
rate for all the \gasLowfb\ (left), \gasHighfb\ (middle), and \gasHighdfb\
(right) simulations.  Here, the no-feedback simulations are shown in blue and
the feedback simulations are shown in green.  The average rate of increase is
calculated from a linear fit of the SFR from $t$=0 to three different times.
The red points are the corresponding points from the \highRes\ runs.  The
linear fit has been performed on the smoothed SFR data. The SFR data are
smoothed as in Figure~\ref{fig:Number_Clump}.  The dark red-dashed straight
lines correspond to visual fits with a slope 1/2 to guide the eye.  One can see
that for higher gas fraction simulations, the SFR growth rate increases with
external pressure and follows the square root of external pressure. At later
times, this is only true up until a certain external pressure. For the low gas
fraction simulation, the SFR growth rate does not scale with the square root of
the external pressure.} 
\label{fig:starFormationSlope}
\end{figure*} 

We expect the  SFR to scale as the square root of the external pressure, at least
in the \gasHigh\ case, for the following reason.  The \cite{Kennicutt1998} star
formation relation is generically fit by
\begin{equation}
\dot \Sigma_\ast={\epsilon\, \Sigma_{g} \over t_{\rm dyn}},
\label{eq:KS1} 
\end{equation} 
for theoretical and observational reasons \citep{E97,S97,Genzel+10}.  In
equation~(\ref{eq:KS1}), 
$\Sigma_\ast$ is the surface density of star formation,
$\Sigma_{\rm g}$ is the gas surface density,
$\epsilon$ is a dimensionless normalisation constant,
and 
$t_{\rm dyn}$ is the dynamical time (the rotation time for a disk galaxy
or the free-fall time for a giant molecular cloud, both classes of objects
fitting the correlation \citealp{KDM12}). The slope of the correlation
corresponds to the linear theory-inspired convolution of gas density and
the most rapidly growing gravitational instability rate for a cold disk.
Much more physics resides in the normalisation, $\epsilon$,
which is a measure of the
star formation efficiency, often defined as the fraction of gas turned into
stars per dynamical time, and in the dispersion. We will not address the
dispersion here, other than to remark that physics beyond a density
threshold must be included, as is evident from the low (e.g. central
molecular zone \citealp{Kruijssen+2014}) and high (both in nearby, cf.
\citealp{Leroy+2015, Turner+2015}  and distant, cf.
\citealp{Finkelstein+2015, Dye+2015}) outliers. 
The usual fit  to the
normalisation is $\epsilon\approx 0.02,$ to star-forming systems out to $z
\sim 2,$ although there is recent evidence that the efficiency is
significantly higher in starburst galaxies at higher redshift.  The logical
generalisation to go beyond a density threshold is to include turbulence
\citep{Padoan+Nordlund2011, HC11, Hopkins+2013}.  An  especially simple
implementation is given in \cite{Silk2001,Silk+Norman2009}.  We will now
extend their argumentation of momentum injection by SNe to an injection of
energy by any kind of process that affects the whole galaxy (e.g. energy
injection given by external pressure).  

Let $E_{{\rm inj}}$ be the kinetic energy injection given for instance  by a
pressure wave coming into the galaxy.  In the same fashion let
$m_{{\rm inj}} v_{{\rm inj}}$ be the momentum injection.  We now make the
educated assumption that the energy injection affects the interstellar clouds.
These clouds acquire terminal velocity given by
\begin{equation}
\sigma _{\rm g} = \epsilon\, \nu_{{\rm inj}} \,{t_{\rm coll} \over t_{\rm dyn}} \ ,
\label{eq:TV1}
\end{equation}
where $t_{\rm coll}$ is the cloud collision timescale and $\nu_{{\rm inj}}$ is the
specific momentum injected.  We immediately see that
the key efficiency parameter $\epsilon$ is proportional to the gas turbulence
velocity $\sigma_g.$ There are recent indications that star formation
efficiency increases in highly turbulent environments \citep{Leroy+2015}.  With
momentum balance and Eq.~(\ref{eq:TV1}), we get 
\begin{equation}
{\dot\Sigma_\ast \,E_{\rm inj} \over m_{\rm inj}\,v_{\rm c}}=
{f_{\rm c}\, {\Sigma_{\rm g}\, \sigma_{\rm g}} \over {t_{\rm coll}}}\ ,
\end{equation}
where $\sigma _{\rm g}$ is the gas velocity dispersion, $v_{\rm c}$ the cloud
velocity and $f_{\rm c}$ the cloud volume filling factor.  We however ignore
complications with $f_{\rm c}$ and define it directly by the cold gas fraction
$f_{\rm g}.$ Also, we assumed $v_{{\rm inj}} \propto v_{\rm c}$, with
$v_{\rm c}$ being the velocity of the cloud.  
We can rewrite the star formation rate per unit volume  as
\begin{equation} 
\dot\rho_\ast=\epsilon_{\rm inj}\,f_g  \sqrt{G\,\rho_{\rm g}}\,\rho_{\rm g}
\label{eq:rhoSFR1}
\end{equation} 
with 
$\epsilon_{{\rm inj}}=({m_{{\rm inj}}v_{\rm c}\sigma_{\rm g}})/{E_{{\rm inj}}}$. 
 
\begin{figure*}
\plotoneF{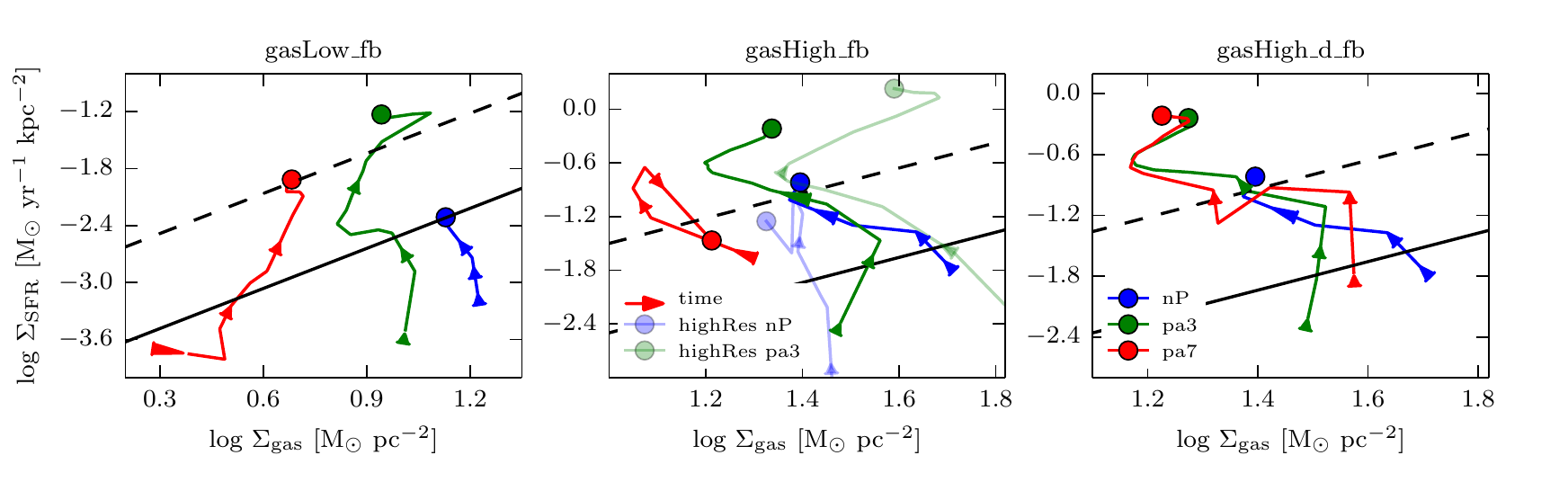}
\caption{Kennicutt-Schmidt (KS) relation for selected runs (nP, pa3, pa7 with
blue, green and red curves, respectively, with lighter colours for the runs
with 4 times higher spatial resolution) from the \gasLowfb\ (left), \gasHighfb\
(middle) and \gasHighdfb\ (right) simulations at 16 different times in color.
The evolution in time is shown with arrows.  The 16 different times are equally
spaced between $t_{\rm start}=25$~Myr and $t_{\rm end}=400$~Myr.  The circle
shown corresponds to $t_{\rm end}$ of the simulation with the corresponding
color.  Data from \citet{Daddi+2010} for the starburst (dashed line) and
quiescent (solid line) sequences are over-plotted for reference. For all the
simulations, independently of the gas fraction, the simulations with external
pressure lie closer to the starburst sequence than the no-pressure simulations.
This shows that our toy model for AGN-induced star-formation might be a
possible explanation for explaining the increased number of starburst galaxies
observed in the distant Universe.}   
\label{fig:KS}
\end{figure*} 

For the present purpose, we note that the SFR can be rewritten as
\begin{equation}
\dot\rho_\ast ={\epsilon_{\rm inj}\over \sigma_{\rm g}}\,G^{1/2}\,\rho_{\rm g}\, p_{\rm
  turb}^{1/2} \ ,
\end{equation}
where we used $p_{\rm g} = \rho_{\rm g}\, \sigma_{\rm g}^2$ and assumed that
$p_{\rm g} \propto p_{\rm turb},$ with $p_{\rm turb}$ being the turbulent
pressure induced by the injection. 

Hence, we expect that AGN-induced pressure should provide a boost  of the star
formation rate, independently of any possible increase in star formation
efficiency, and initially vary as the square root of the pressure boost. This
effect should be important for the gas-rich simulations as gas turbulence plays
a lesser role in contributing to a gas-poor disk and the assumption that the
gas pressure scales with the gas turbulence breaks down. This is because we
omitted the cold cloud mass fraction in the simple formulation above, and the
molecular hydrogen fraction decreases at lower gas pressure
\citep{BlitzRosolowsky2006}.
 
Fig.~\ref{fig:starFormationSlope} illustrates how the SFR increases in time as
a function of the external pressure.  The mean SFR is calculated from a linear
fit of the SFR from time $t=0$ to time $t$.  For the \gasLow, \gasHigh, and
\gasHighd\ simulation sets, the SFR increases with increasing external pressure
until a maximum is reached and then decreases again or stays at a similar level
as for the \gasHighd\ simulations.  Also, no significant difference in the SFR
can be seen between the feedback and non-feedback simulations, indicating that
the external pressure increase is the dominant effect driving the increased
SFR.  

In order to compare with the prediction explained above, a dark red dashed
curve is plotted to guide the eye in Fig.~\ref{fig:starFormationSlope}
representing a square root fit of the SFR as a function of the external
pressure applied on the galaxy.  One sees that, for the gas-rich \gasHigh\ and
\gasHighd\ simulations, the SFR follows the square root of the external
pressure very well. At later times, the square root law is only fulfilled until
the optimal pressure is reached. For the \gasLow\ simulations, the SFR does not
scale well with the square root of the external pressure. But as explained
above, this was  not unexpected.

The bright red points in Fig.~\ref{fig:starFormationSlope} are the
corresponding points from the \highRes\ run.  As can be seen in
Appendix~\ref{sec:Convergence} the formation of stars in the \highRes\ run
shows a delay compared to the \lowRes\ run. This is due to the increase in star
formation threshold for the higher resolution run. Because of this delay, one
can also see a delayed behaviour of the SFR of the \highRes\ run in
Fig.~\ref{fig:starFormationSlope}.  However, at later times, the \highRes\
simulation shows a similar behaviour to the \lowRes\ simulation. 
 
\subsection{The Kennicutt-Schmidt Relation}
 
The Kennicutt-Schmidt (KS) law \citep{Kennicutt1998} law relates the SFR per
unit area as a power of the surface density of gas. This relation holds over
several orders of magnitude in both quantities (\citealp*{KDM12}, and
references therein), with the same normalisation for global galaxies (including
high redshift ones, \citealp{Genzel+10}) and giant molecular clouds in the
Milky Way (\citealp{Heiderman+10}; \citealp*{LLA10}) and in the nearby M51
galaxy \citep{Kennicutt+07}.  This demonstrates the remarkable universality of
the SFR.  At high redshift, starburst galaxies lie above the KS law for normal
galaxies \citep{Genzel+10}.  While the cause of this observed offset is not
known, one may speculate that this increased SFR may be caused by
positive AGN feedback.

To investigate this in further detail, we check whether our pressurised
galactic discs follow the KS relation for normal galaxies, or are above it as
starbursts are observed to be, or below it.  We adapt here an equivalent
technique to \cite{Powell+2013} and calculate the half-light radius by
assigning a luminosity to each star particle dependent on their age and
proportional to their mass \citep{Weidner+2004},
\begin{eqnarray}
L(\rm age < 10\,{Myr}) &\propto& M_{\rm stars} \\
L(\rm age > 10\,{Myr}) &\propto& M_{\rm stars}\rm \left( \frac{age}{10\,{Myr}} \right)^{-0.7}
\end{eqnarray} 
We randomly assign an age in the range 0--5~Gyr for stars that are specified
in the initial conditions and therefore have an age equal to zero when the
simulation starts. For a given output, $\Sigma_{\rm SFR}$ is calculated within
the half-light radius using the SFR averaged over the previous 10~Myr.  The KS
relation is calculated by finding the 3D half-light-radius.
Within this volume, all the gas above a threshold of $0.1\, \rm H\,cm^{-3}$ is
used to calculate $\Sigma_{\rm{gas}}$ and all the new stars are used to
calculate $\Sigma_{\rm SFR}$, however the quantities are divided by the area
$\pi r_{\rm{3D}} ^ 2$. 

Fig.~\ref{fig:KS} shows the KS relation at different times for several
simulations for the three cases of \gasLowfb, \gasHighfb, and \gasHighdfb\ with
the observed relation from~\cite{Daddi+2010} overplotted.  One can see that,
over the course of 400 Myr, all runs lead to an increase in $\Sigma_{\rm SFR}$,
by one to two dex, with much smaller variations (less than 0.3 dex) in
$\Sigma_{\rm gas}$.  In particular, the runs with high gas fraction (with or
without external pressure) show a decrease in $\Sigma_{\rm gas}$.  This
decrease in gas surface density is related to the gas mass outflow at late
times (Fig.~\ref{fig:MassOutflow}) and to the consumption of gas by star
formation.  For the \gasLow\ simulation, the evolution of $\Sigma_{\rm gas}$ is
tied to the evolution of the total baryonic mass within the galaxy shown in
Fig.~\ref{fig:totalMass}.  The light blue and green lines in Fig.~\ref{fig:KS}
show the \highRes\ simulations for the nP and pa3 set, respectively. One can
see that the trends of the \highRes\ simulation are very similar to the trends
of the \lowRes\ simulations,  specifically at late times. Especially for the pa3
\highRes\ simulation, one can see that higher gas densities are reached due to
the higher resolution  which allows  the gas  to collapse even further.

For all the simulations independently of the gas fraction, the simulations with
external pressure end up being pushed closer to or further beyond  the
starburst sequence than the corresponding simulations without external
pressure. This trend is not changed for the \highRes\ runs.  Runs with external
pressures leading to  higher SFR also have a higher $\Sigma_{\rm SFR}$ and
therefore end up even closer to the starburst sequence.  For instance, for the
\pspher\ simulations, the pa7 run does not extend as far beyond the observed
starburst sequence as the pa3 run, which indeed reaches higher SFR for both the
\gasLowfb\ and \gasHighfb\ simulations (Figs.~\ref{fig:SFR} and
\ref{fig:starFormationSlope}).  This is not as significant for the \gasHighdfb\
simulations as they do not experience a decrease of SFR but rather that the SFR
stays at a certain level after a certain pressure increase ($\sim$~pa5). We
therefore see for this simulation set that the KS relations end in a very
similar parameter space.

We conclude that this toy model for AGN-induced over-pressurisation plausibly leads to
AGN-associated star-forming galaxies having enhanced specfic star formation
rates, for example as  suggested by recent observations, cf.
\citep{Zinn+2013, Drouart+14}.  
 
\section{Conclusions}
\label{sec:conclusions}
 
It is a fascinating challenge to understand the extreme star formation rates
observed for some high-redshift galaxies,
typically with luminous AGN and massive outflows: are these caused by higher
contents of molecular gas or by a greater efficiency of star formation
relative to this molecular gas content? Is turbulence sufficient to explain
the high SFR values, or do we need
recourse to a more exotic pathway that enhances star formation rates even more? The
latter option is motivated by the increasing evidence for the role of AGN in
star formation, and in particular their role in a putative phase of positive
feedback that accompanies or even precedes the commonly observed massive, star
formation-quenching, outflows stimulated by AGN activity.  

Using hydrodynamical simulations of isolated disc galaxies embedded in a hot
over-pressurised halo, we have been able to study the response of the galaxy
SFR to the forcing exerted by this external gas pressure onto the disc.  The
pressure enhancement triggers instabilities leading to more fragmentation when
compared to the no-pressure simulations (Figs.~\ref{fig:Map_bulge1_ad_fb} and
\ref{fig:Map_comp}). The enhanced fragmentation leads to the formation of more
clumps (Fig.~\ref{fig:Number_Clump}) as well as larger values of SFR
(Fig.~\ref{fig:SFR}). This hints at a positive effect of the pressurisation of
the disc and therefore to positive feedback. 

We observe a difference in the behaviour for the different ways in which the pressure is
applied. In the simulations where external pressure is continuously
applied beyond a certain
radius (\pspher\ simulations), we observe an optimal pressure
beyond which the number of clumps as well as the SFR is decreased. For the
simulations where the pressure is instantaneously applied using a density threshold
(over-pressure applied closer to the galaxy disc), such an optimal pressure
is not observed. 

We have seen that the mass outflow plays a role in explaining
this optimal pressure. In particular, for the \gasHighfb\ simulations, a
significant amount of gas gets expelled out of the galaxy, leaving little gas
left to form stars and thereby lowering the SFR. The difference in SFR
between the high and low external pressures for the
\gasLowfb\ simulations is explained by the stagnation of the accumulation of
mass in the clumps, which is again related to the large amount of gas that is
removed by the incoming pressure wave.  Our simulations have been tested with respect to the
resolution and local presence or absence of SN explosions: the
over-pressurisation of the disc still leads to a positive feedback effect
(enhanced SFR).

We found that at given times of the \pspher\ simulations, the SFR (and its mean
growth rate) vary as the square root of the applied pressure.  We explain this
by adapting the Schmidt law for the SFR as a function of 3D gas density for the
inclusion of extra pressure
caused by the AGN bow shock-driven radio lobe or wind, leading to compression
times typically an order of magnitude shorter than the dynamical time, as
argued by \cite{Silk+Norman2009}. 

Though our setup of the extra pressure exerted by circumgalactic gas onto the
galaxy is crudely modeled to mimic the pressure confinement by  AGN activity,
we are confident that such a mechanism could operate in more realistic
configurations (see the jet simulations of~\citealp{Gaibler+2012}).  We have
demonstrated that such pressure confinement of the ISM drives the galaxy into
an intense star formation regime, and could explain observations of star
formation-enhanced galaxies in the presence of jet activity~\citep{Zinn+2013}.
Cosmological simulations of pure AGN jet feedback in galaxy
clusters~\citep{duboisetal2010} have shown that it has a negative impact on the
galaxy SFR on the long-term, though these simulations were lacking  spatial
resolution in order to properly capture the small-scale fragmentation of the
ISM.  Our more global picture could suggest a two-stage mechanism for AGN
feedback: a compression phase leading to a short burst of star formation,
together with the expulsion or heating of the circumgalactic gas leading to a
suppression of the gas accretion onto the galaxy and its star formation on
longer time-scales.  This remains to be verified with simulations of galaxies
embedded in a cosmological environment with high spatial resolution and a
self-consistent treatment of AGN feedback.  We defer this study to future work.

\section*{Acknowledgments}
YD and JS acknowledge support
by  ERC project 267117 (DARK) hosted
by UPMC -- Sorbonne Universit\'es and JS for support at JHU by National Science
Foundation grant OIA-1124403 and by the Templeton Foundation.  RB has been
supported in part by the Balzan foundation and the  Institute Lagrange de
Paris.  This work has been partially supported by grant Spin(e) ANR-13-BS05-
0005 of the French ANR.  The simulations have made use of the Horizon cluster.
We specially thank S. Rouberol for technical support with the horizon cluster
at IAP.  We also thank M. D. Lehnert, M. Volonteri, A. Wagner, and J. Coles for
valuable discussions. 

\bibliographystyle{mn2e}
\bibliography{refs}

\appendix

\section{Bipolar pressure increase}
\label{sec:Bipolar}
\begin{figure}
 \centering
 \leavevmode
 \includegraphics{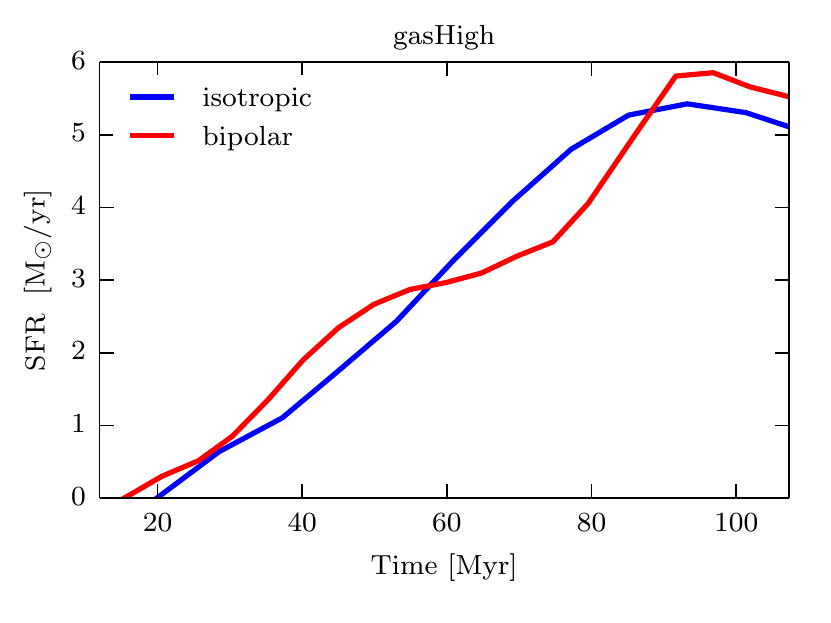}
\caption{Star formation rate (SFR) as a function of time: In blue for the case
where the pressure is applied isotropically (\textit{isotropic}) and in red
when the pressure is applied to the galaxy in a bipolar geometry.
(\textit{bipolar}).  }
\label{fig:Bipolar}
\end{figure}

%
To study the assumption of a isotropic pressure increase, we have performed a
simulation of a non-isotropic bipolar pressure increase.  For this the pressure
has only been increased after a certain height (1.5~kcp) in the vertical
direction of the galaxy, were the pressure has been kept at the normal value in
the radial direction.   The SFR of the \textit{bipolar} and \textit{isotropic}
simulations are shown in Fig.~\ref{fig:Bipolar}.  One can see that while the
\textit{bipolar} SFR oscillates more the general behaviour is not changed by
the way pressure is applied on the galaxy.  
 
\section{Effects of supernova feedback}
\label{sec:SNnSN}
\begin{figure}
\plotone{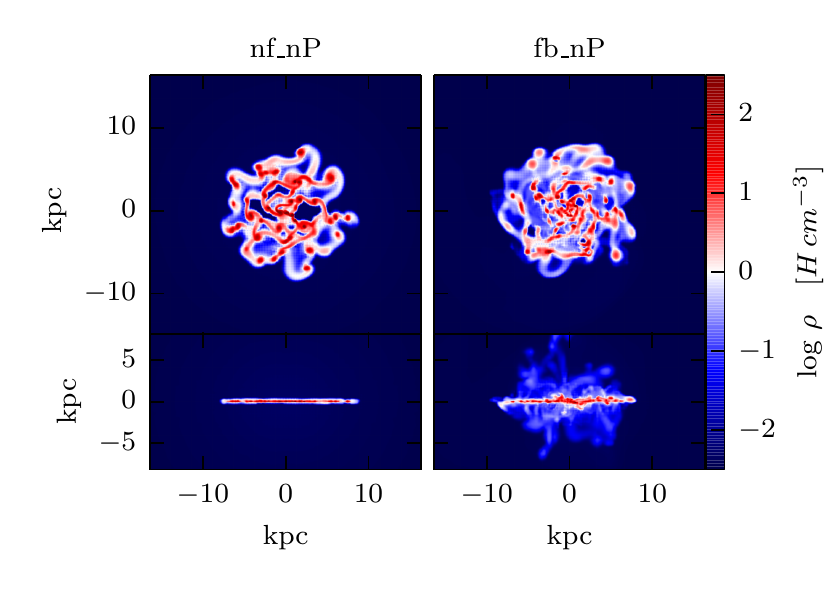}
\caption{Gas density maps (mass-weighted) of the \gasHigh\ non-feedback (left) and feedback (right) simulations
without enhancement of the external pressure (nP). The maps are taken at the end
of the simulation ($\sim 400$~Myr). Each panel shows both face-on (40$\times$40 kpc, upper part) and
edge-on (40$\times$20 kpc, lower part) views.} 
\label{fig:MapComp1}
\end{figure}

\begin{figure}
\plotone{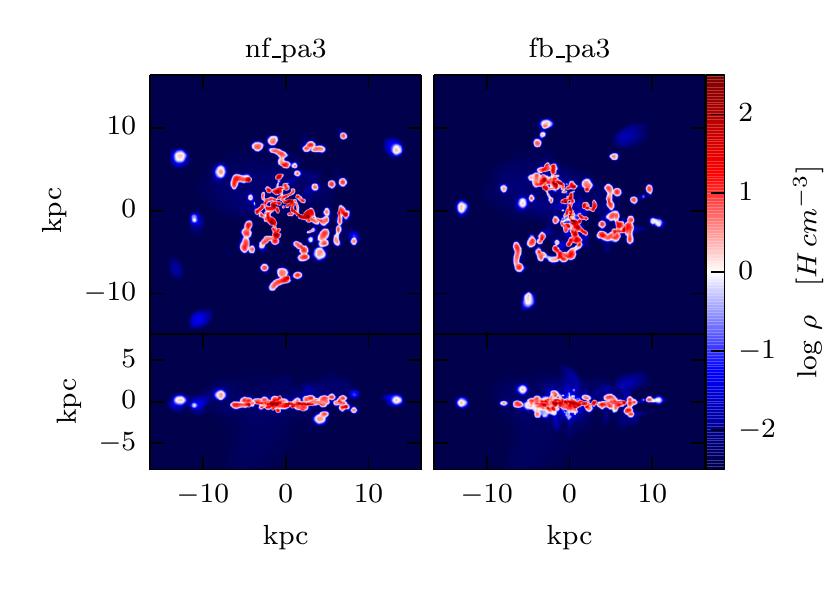}
\caption{Similar as Fig.~\ref{fig:MapComp1} but for the simulations with pressure
enhancement pa3.} 
\label{fig:MapComp2}
\end{figure}

Here, we compare the feedback run with the no-feedback run.  In
Fig.~\ref{fig:MapComp1} the gas density maps of the no-pressure enhancement
simulations are shown for the non-feedback (nf, left panel) and feedback (fb,
right panel) simulations.  In Fig.~\ref{fig:MapComp2} the comparison between fb
and nf is shown for the pa3 simulations.  We see that for the no-pressure
simulations, the effect of SN explosions is to disrupt the interstellar medium
into smaller but more numerous clumps. In the edge-on-view, we can also see that the 
feedback simulation thickens the disc and enhances the mass outflow close to the
galaxy. For the pressure simulation, no significant difference can be observed.
It shows that the effect of external pressure is stronger than the effect
of SN explosions. 

\begin{figure*}
\plotoneF{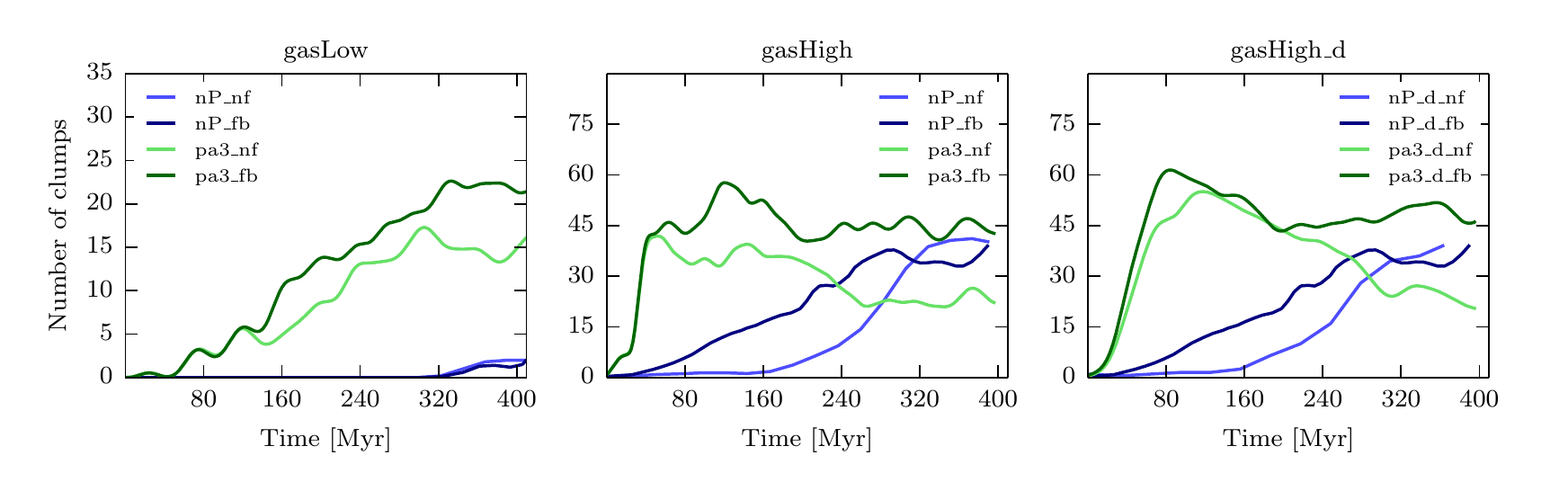}
\caption{Time evolution of the number of clumps for a selection of the \gasLow\
(left), \gasHigh\ (middle), and \gasHighd\ (right) \gasHigh\ simulations. For each
simulation set the feedback (fb) and non-feedback (nf) runs are shown for
comparison. They are indicated by the suffixes in the legend. The lines
are smoothed as in Figure~\ref{fig:Number_Clump}.}
\label{fig:NumberClumpComp}
\end{figure*}

\begin{figure*}
\plotoneF{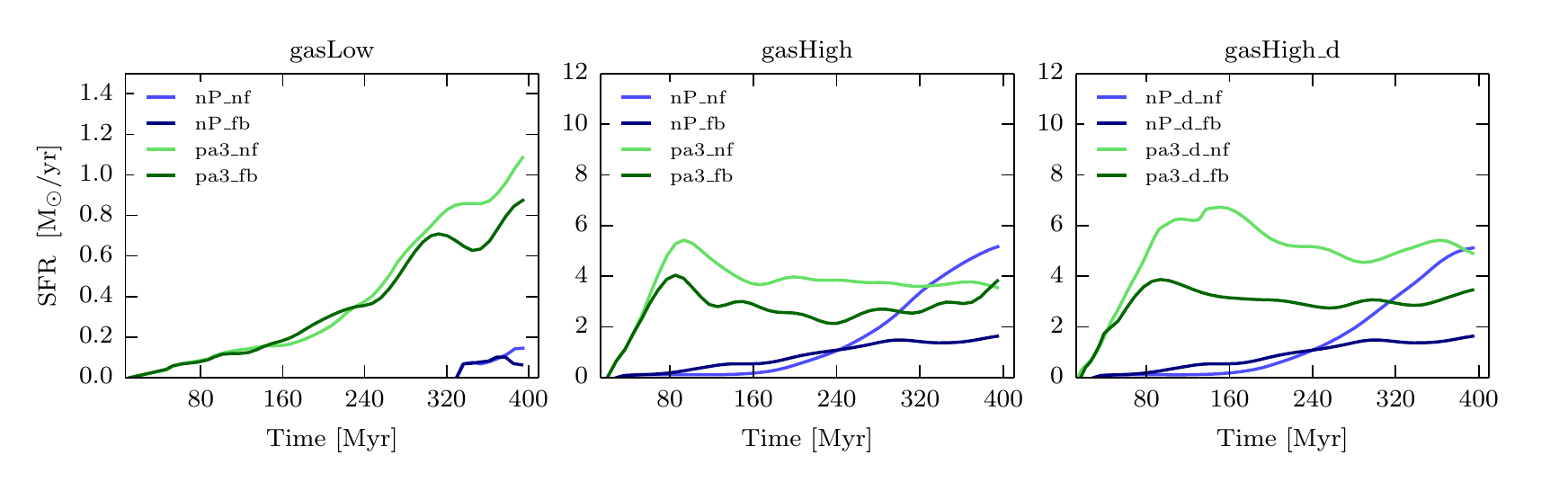}
\caption{Time evolution of SFR for a selection of the \gasLow\ (left),
\gasHigh\ (middle), and \gasHighd\ (right) simulations. For each simulation
set the feedback (fb) and non-feedback (nf) runs are shown for comparison.
They are indicated by  suffixes in the legend.  The lines are
smoothed as in Figure~\ref{fig:Number_Clump}.}
\label{fig:SFRComp}
\end{figure*}

In Fig.~\ref{fig:NumberClumpComp}, we show the number of clumps as a function of
time for a selection of the \gasLow\ (left), \gasHigh\ (middle), and \gasHighd\
(right) simulations with (fb) or without (nf) SN feedback.  In
Fig.~\ref{fig:SFRComp}, we show the time evolution of the SFR for the same
selection of runs.  We see that the number of clumps is enhanced by the
presence of SN explosions in all cases since the clumps are regularly destroyed by
the SN activity~\citep{duboisetal15}.  SNe regulate the mass growth of the gas
clumps, and since the most massive clumps are expected to capture the smaller
clumps, SNe allow for the increase in the number of clumps, thereby reducing
their average cross section and mass (see Fig.~\ref{fig:ClumpPropMass}).  We also see that the
SFR is higher for the non-feedback simulation compared to the feedback
simulations as a consequence of the absence of a local regulating process
within gas clumps.

Reassuringly, the effect of over-pressurisation of the disc onto the SFR
enhancement is independent of the presence of SN explosions: it still leads to
a positive feedback effect that SNe only marginally modulate.

\section{Convergence Studies}
\label{sec:Convergence}

In this section, we test how the results depend on the resolution of the
simulation. We performed two high resolution (\highRes) simulations for the
\gasHigh\ case, one with no external pressure (nP\_hR) and the other with
external pressure (pa3\_hR). The higher resolution runs have been performed
with a resolution of  $\Delta x = 10\,\rm kpc$ (compared to 40 kpc for the
standard runs).   We changed the density threshold for star formation
($n_{0}=224\,\rm H\, cm^{-3}$) in the polytropic EoS as well as the dissipation
time-scale of the non-thermal component for the SN feedback ($\Delta x=10\,\rm
pc$) with the resolution. The simulations were run for a similar timescale
($\sim 400 \,\rm Myr$) as the lower resolution (\lowRes) simulations. 

\begin{figure}
\plotone{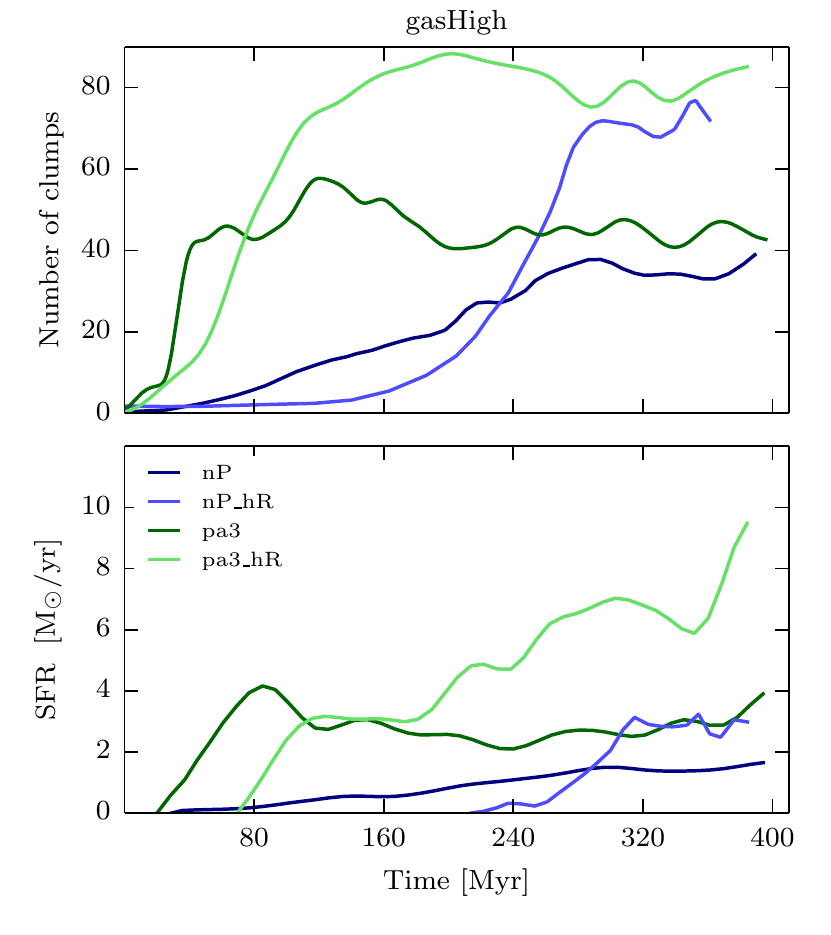}
\caption{Time evolution of the number of clumps (upper panel) and SFR (lower
panel) for the low resolution and high resolution \gasHighfb\ simulations. The 
clumps were extracted using a gas density threshold of $21 \,\rm H\, cm^{-3}$. }
\label{fig:CompHRLR}
\end{figure}

In Fig.~\ref{fig:CompHRLR}, we show the comparison between the \highRes\ and
\lowRes\ simulations. In the upper panel, the number of clumps is shown for the
high and low resolution runs where for both simulations the same clump
detection density threshold of $21\,\rm H\, cm^{-3}$ and a peak-to-saddle
threshold of 1.5 was chosen. 

Fig.~\ref{fig:CompHRLR} shows that, in both \highRes\ and \lowRes\ runs, clumps
are formed at a faster rate when over-pressure is applied on the galaxy.
Comparing the two resolution runs, we see that the rates of clump formation for
both resolutions are comparable at the start of the simulations, for both the
pressure and no-pressure runs. While the \lowRes\ run with external pressure
(pa3) sees a sharp rise in its clump number at 25 Myr, the number of clumps in
the \highRes\ run with external pressure (pa3\_hR) starts catching up after 50
Myr and soon (at 70 Myr) overtakes that of the pa3 run, to end up with nearly
double the number of clumps.  A similar effect is seen in the no-pressure runs:
the number of clumps in the \highRes\ simulation starts slowly, but overtakes
that of the \lowRes\ run (at 230~Myr) to also end up with nearly double the
number of clumps.

Similar trends are seen in the star formation histories (lower panel of
Fig.~\ref{fig:CompHRLR}).  For the no-pressure runs, the \highRes\ one
overtakes the other one in SFR at 280 Myr to end up with twice the SFR, while
in the corresponding runs with external pressure, the \highRes\ one has its SFR
overtake that of the \lowRes\ analog at 150 Myr, end the \highRes\ run ends up
with over double the SFR of the \lowRes\ one.  The very slow rise of the SFRs
in the \highRes\ runs is the consequence of our choice of a higher density
threshold for the \highRes\ simulations, which is reached at later times.  Once
stars start to form, the SFR is greater in the pressure simulation than in the
no-pressure simulation. The general effect that the pressurisation leads to
more star formation is therefore still the same.

\end{document}